\documentclass[11pt,a4paper]{article}
\pdfoutput=1
\usepackage{jheppub}

\def\numu{{\nu_{\mu}}}

\newcommand{\eg}{{\it e.g.}}
\newcommand{\ie}{{\it i.e.}}

\newcommand{\beq}{\begin{equation}}
\newcommand{\eeq}{\end{equation}}
\newcommand{\beqa}{\begin{eqnarray}}
\newcommand{\eeqa}{\end{eqnarray}}

\usepackage{cancel}
\usepackage{soul}

\usepackage{paralist}
\makeatletter
\renewcommand\p@enumii{}
\makeatother
\newcommand\Tstrut{\rule{0pt}{2.6ex}}         

\title{Requirements on common solutions to  the
 LSND and MiniBooNE excesses: a post-MicroBooNE study}
\author[a]{Waleed Abdallah,}
\author[b]{Raj Gandhi}
\author[c]{and Samiran Roy}
\affiliation[a]{Department of Mathematics, Faculty of Science, Cairo University, Giza 12613, Egypt}
\affiliation[b]{Harish-Chandra Research Institute (A CI of the Homi Bhabha National Institute), Chhatnag Road, Jhunsi, Allahabad 211019, India}
\affiliation[c]{Institute of Physics, Sachivalaya Marg, Sainik School Post, Bhubaneswar 751005, India}
\emailAdd{awaleed@sci.cu.edu.eg} 
\emailAdd{raj@hri.res.in}
\emailAdd{samiran.roy@iopb.res.in}

\abstract{
The strong statistical significance of  an observed electron-like event excess in the MiniBooNE (MB)  experiment, along with an earlier similar excess seen in the Liquid Scintillator Neutrino Detector (LSND), when interpreted in conjunction with recent MicroBooNE results may have brought us to the cusp of new physics discoveries. This  has led to many attempts to understand these observations, both for each experiment individually and in conjunction,  via physics beyond the Standard Model (SM). We  provide an overview of the current situation, and discuss three major categories under which the many proposals for new physics fall.  The  possibility that the same new, non-oscillation  physics explains both anomalies leads to new restrictions and requirements. An important class of such common solutions, which we focus on in this work,  consists of a heavy ${\cal O}$(MeV$-$sub-GeV) sterile neutral fermion produced in the detectors,  (via up-scattering of the incoming muon neutrinos), and subsequently decaying to photons or $e^+e^-$ pairs which mimic the observed signals. Such solutions are subject to strong  demands from  a) cross section requirements which would yield a sufficient number of total events in both LSND and  MB, b) requirements imposed by the measured energy and angular distributions in both experiments and finally, c) consistency and compatibility of the new physics model and its particle content with other bounds from a diverse swathe of particle physics experiments. We find that these criteria often pull proposed solutions in different directions, and stringently limit  the viable set of proposals which could resolve both anomalies. Our conclusions are relevant for both the general search for new physics and for the ongoing observations and analyses of the MicroBooNE experiment.
}

\begin{document}
\maketitle
\section{Introduction}
\label{sec1}
The observation of unexplained electron-like excesses in  the Liquid Scintillator Neutrino Detector (LSND)~\cite{LSND:1996vlr} and the MiniBooNE (MB) experiment~\cite{AguilarArevalo:2008rc, Aguilar-Arevalo:2018gpe,MiniBooNE:2020pnu} has provided an important window of opportunity  in our search for  physics beyond the Standard Model (SM). This importance is based on the strong statistical significance of these excesses ($3.8\sigma$ for LSND~\cite{LSND:2001aii} and $4.8\sigma$ for MB~\cite{MiniBooNE:2020pnu}, with a combined significance of $6.1\sigma$~\cite{Aguilar-Arevalo:2018gpe}).  The results are  buttressed by  careful estimates and measurements  of possible SM backgrounds in order to eliminate standard physics explanations. (For discussions, see \eg~\cite{LSND:1996jxj,Katori:2020tvv,Dasgupta:2021ies,Brdar:2021cgb,Alvarez-Ruso:2021dna,MicroBooNE:2021zai} and references therein.) 

Both experiments differed in beam energy by an order of magnitude, and had very different systematic uncertainties and backgrounds, implying that if indeed standard physics was responsible for the excesses observed, it would likely require two qualitatively different  explanations. While MB confirmed the presence of an electron-like excess in the low-energy   region of its range in its effort to validate the LSND observations, both LSND and MB were insensitive to whether the excess originated in electrons, photons or collimated $e^+e^-$ pairs\footnote{In what follows, we use the acronym ``LEE" (for low energy excess) to refer to both the LSND and the MB excesses, although in the literature for the most part it is used for the MB excess only.}.  The MB neutrino fluxes peak around  800 MeV while the energy of  the primary decay-at-rest (DAR) flux ($\bar{\nu}_{\mu}$) of LSND was significantly lower, going down to as low as $\sim 50$~MeV. LSND also had a decay-in-flight (DIF) flux, going  up to $\sim 250$~MeV.  

Very recently, the MicroBooNE experiment~\cite{MicroBooNE:2016pwy}, designed specifically to be sensitive to differences between the various signal possibilities mentioned above and help pin down the origin of these anomalies, announced  its first set of results~\cite{MicroBooNE:2021zai,MicroBooNE:2021nxr,MicroBooNE:2021rmx, MicroBooNE:2021sne,MicroBooNE:2021jwr}, which seem to offer further support for the presence of new physics. A major goal of the  experiment is to search for an LEE signal in $\nu_e$ charged-current (CC) interactions with an electron in the final state  (the eLEE search) as well as in neutral-current (NC) interactions with a single photon in the final state. Four separate searches and analyses were carried out: \begin{inparaenum}[i)]\item\label{uB:A1} a final state photon produced by $\Delta$ production and decay; \item\label{uB:A2} an exclusive search using the charge current quasi-elastic (CCQE) interaction channel; \item\label{uB:A3} a semi-inclusive search for events with one electron and no final state pions; and \item\label{uB:A4} a fully inclusive search for events with one electron and any other possible final state. \end{inparaenum}
MicroBooNE expected, but did not see, any excess in electrons beyond SM backgrounds.  Their results thus so far indicate that the MB excess was not due to electrons. Moreover, the most likely photon-producing SM background  ( \ref{uB:A1}) above) was measured to be consistent with expectations. 

We note that these results are based on a part ($6.8  \times 10^{20}$ protons on target) of what will eventually comprise the full MicroBooNE data sample, and may change with further data and analysis. However, in spite of the fact that MicroBooNE saw results in conformity with the SM, at the present time these seem to point (albeit not unequivocally) to the possibility that the excess could stem from a) single photons or b) the production of $e^+e^-$ pairs, with their origin lying in new physics. This conclusion is based upon the following reasoning:
\begin{itemize}
\item It is assumed that the MB LEE is {\it{robust}}. This assumption is based on  $(a)$ data-driven background measurements, $(b)$ comparisons with state-of-the-art neutrino generators and $(c)$ estimates of theoretical uncertainties, as discussed in~\cite{Katori:2020tvv,Brdar:2021cgb} and references therein.
\item By confirming that the largest and most likely single photon background is in conformity with production rates dictated by the SM~\cite{MicroBooNE:2021zai}, MicroBooNE has pointed out that if the LEE is due to photons, they are likely produced via new physics arising from mechanisms like radiative decay of heavy neutrals and/or non-standard magnetic moments. This conclusion is to be drawn in juxtaposition with MB’s own measurements of control samples of single photons from $\pi_0$ production and decay, and from $\Delta$ (1232) production and decay~\cite{AguilarArevalo:2009ww}.
\item As mentioned above, via three searches for $\nu_e$ scattering leading to the production of electrons \cite{MicroBooNE:2021nxr, MicroBooNE:2021jwr, MicroBooNE:2021rmx, MicroBooNE:2021sne}, each with a different final state, MicroBooNE excluded an electron final state parametrisation of the MB excess (the eLEE model) at a confidence level $> 97\%$. By doing so it disfavoured (but did not eliminate) the possibility that the excess originated in sterile-active neutrino oscillations and strengthened the possibility that the excess could be due to boosted $e^+e^-$ pairs, which, not having a known SM background, points to new physics.
\end{itemize}

Given the potential for discovery, it is important to examine  aspects of the possible new physics scenarios and  their implications. We attempt to do this by first  providing an overview of the present situation as it stands post the MicroBooNE results in section~\ref{sec2}. The rest of our paper is focussed on studying an important class of new physics solutions which will soon be tested by MicroBooNE and subsequent experiments.  Specifically, we study, from a  phenomenological perspective, the  demands  placed on non-oscillation new physics which attempts to explain  both these anomalies simultaneously.  We restrict ourselves to a class of solutions where a new heavy sterile neutrino is produced inside either detector (LSND or MB), via up-scattering of the incoming beam $\nu_\mu$ (or $\bar{\nu}_\mu$), and which subsequently decays in a manner that mimics the observed signal for the excess by producing an $e^+e^-$ pair or a photon.  This final state is produced by new fermions and/or boson mediators~\cite{Bertuzzo:2018itn,1808.02915,1903.07589,Dutta:2020scq,Abdallah:2020biq,Abdullahi:2020nyr,Abdallah:2020vgg,0902.3802,1009.5536,Fischer:2019fbw,Datta:2020auq,2110.11944,2105.06470}. We find that once one fits MB, important restrictions ensue on demanding that the same benchmark parameters also explain LSND, narrowing and illuminating the search for new physics. The conclusions  we obtain are also relevant to observations and analyses focussed on new physics currently in progress in the MicroBooNE experiment.

Section~\ref{sec3} describes the  up-scattering interaction in the detector and provides expressions for the cross section and event rate. Section~\ref{sec4} focusses on the variation of the cross section with incoming neutrino energy for two different types of mediators (neutral vectors and CP-even scalars)\footnote{Hereafter, we simply refer to these two mediators as vector and scalar mediators.} and also examines the behaviour of the cross section with different mediator masses. In section~\ref{sec5} we study the requirements imposed by fitting the MB angular distribution. Section~\ref{sec6} examines information that can be obtained by {studying} the energy distributions in both detectors, while section~\ref{sec7} studies the  kinematics of decay of the up-scattered heavy neutrino and its effect on the requirements. Section~\ref{sec8} collates the information in sections~\ref{sec2}$-$\ref{sec7} to arrive at a set of final requirements on a common solution. Finally, section~\ref{sec9} summarizes our conclusions. 
\section{The LEE and its resolution via new physics: a post-MicroBooNE overview}
\label{sec2}
If one were to classify the many new physics proposals\footnote{These include proposals that attempt to explain one or the other anomaly, or both simultaneously.} made to understand the LEE keeping the recent  MicroBooNE results in mind,  one finds that most  of these fall under one of  three broad categories~{\bf\ref{Prop-a})},~{\bf\ref{Prop-b})}~or~{\bf\ref{Prop-c})} below, depending on the final state they produce in these detectors\footnote{Some proposals cannot be classified easily within the simple scheme we have chosen here, since they combine two of the three final states, \eg~\cite{Datta:2020auq,2105.06470,2110.11944}.}. This classification is convenient because of MicroBooNE's ability to distinguish among them:
\begin{enumerate}[\bf a)]
\item\label{Prop-a} Proposed solutions  in this category are distinguished by the fact that their final state, both in LSND and MB is a {\it{single electron or positron}}. Since the announcements of the LSND and subsequently, the MB results,  much attention over the past decade or more has been  focussed on light sterile neutrinos of mass ${\cal O}$(eV), with consequent active-sterile mixing and oscillations to $\nu_e$ or $\bar{\nu}_e$ to account for the observed signals. We note that in addition to  LSND and MB,  other short baseline experiments at nuclear reactors~\cite{1101.2755},  and radioactive source experiments~\cite{1006.3244, 2109.11482} have reported anomalous signals involving electron neutrinos. The picture regarding their inter-connectedness to LSND and MB, however, remains unclear. Recent reviews and references on active-sterile oscillation solutions to the anomalies may be found in~\cite{ 2110.09876,Dasgupta:2021ies,2109.13541,1907.08311, 1906.00045,1903.04608}. Variants on the simple vacuum oscillation solution,  which use matter effects and  resonant oscillations with new mediators to produce a single electron or positron in the detector  may be found in~\cite{Alves:2022vgn} (vector  mediator) and~\cite{1712.08019} (scalar mediator).
  
 A second subset of solutions,  which fall into this category because they produce the same final state,  but are distinct from the point of view of the physics involved, invoke the production of additional $\nu_e$ (or $\bar{\nu}_e$) in the beam. These $\nu_e$ have their origin in  the decay of a  heavy  neutrino ($ \nu_4$, eV to hundreds of keV in mass)  to mostly-active neutrinos via $\nu_4 \to \nu_i + \phi$, where $\phi$ is a new scalar and $\nu_i$ an SM neutrino~\cite{hep-ph/0505216,1512.05357,1711.05921,Dentler:2019dhz, deGouvea:2019qre}. 
  
Both subsets of proposals in category~{\bf\ref{Prop-a})} remain in contention due to reasons described in recent work~\cite{2111.05793,2111.10359,Alves:2022vgn,MiniBooNE:2022emn}, $\ie$ that MicroBooNE results disfavour but do not eliminate the  possibility  that LSND and MB results can be explained by i) ${\cal O}$(eV) sterile neutrinos with active-sterile mixing and oscillations, or more generally, by ii) the presence of additional $\nu_e$ in the beam. {Arguments about the  viability of these  choices  draw} upon a reanalysis~\cite{2111.05793} of the MicroBooNE data viewed in conjunction  with other indicators of active-sterile oscillations, like the source experiments (SAGE, GALLEX and BEST~\cite{1006.3244, 2109.11482}) and the reactor flux calculations and data~\cite{1101.2663,1106.0687,1101.2755}\footnote{We note that recent improved calculations~\cite{2110.06820} of inverse beta decay yields at reactors appear to disfavour the existence of a reactor anomaly and point to a significant tension with the  parameter space identified by the Gallium source experiments, making the link between the two tenuous.}. They also assign weightage to the systematic uncertainties present in the backgrounds at MB. Importantly, they rely on the fact that MicroBooNE mapped the MB excess to an eLEE model which is not unique, in that other $\nu_e$ spectral choices can also reproduce the MB excess~\cite{2111.10359,Alves:2022vgn}. These considerations  render MicroBooNE's coverage of MB's parameter space less certain and extensive, and consequently do not rule out these proposals.
\item\label{Prop-b} A second category of proposals proceeds on the basis  that \begin{inparaenum}[i)]\item\label{C-b-i} the conflict  with  $\nu_\mu$ disappearance data (discussed below),  \item\label{C-b-ii} the tension with cosmology (below), \item\label{C-b-iii} the apparent discrepancy between the oscillation parameter space identified by source experiments and reactor data~\cite{2110.06820} along with \item\label{C-b-iv} the recent MicroBooNE results\end{inparaenum}, when considered in conjunction,  constitute sufficient evidence that 
while heavier sterile neutrinos might well be part of nature's particle spectrum, they do not seem to reside in the mass and mixing parameter space singled out by oscillation explanations of the various low energy anomalies discussed above. The final state in  these proposals  is not a single electron or positron, but {\it{an $e^+e^-$ pair}} produced in the detector~\cite{0902.3802,Bertuzzo:2018itn,1808.02915,1903.07589,Fischer:2019fbw,Datta:2020auq,Dutta:2020scq,Abdallah:2020biq,Abdullahi:2020nyr,Abdallah:2020vgg,Chang:2021myh,2105.06470,2110.11944}. The focus  is on finding other  explanations  involving new physics in production and decay for the LSND\footnote{LSND did not conduct a search for $e^+e^-$ pairs or $\gamma\gamma$ pairs.  Due to  this reason it is reasonable to assume that it would reconstruct most $e^+e^-$ pairs as a single electron event. Specifically,  since timing was their most powerful particle identifying variable, the fit to a Cherenkov ring would automatically select the most significant ring, even for large angles between the $e^+e^-$ pair. For this reason,  $e^+e^-$ pairs with correlated neutrons could explain the LSND excess.} and~MB anomalies. Many such proposals postulate a new physics interaction involving a vector or scalar mediator which is not part of the SM spectrum, as well as heavy sterile neutrinos, several orders of magnitude higher in mass than those in the oscillation scenarios of category~{\bf\ref{Prop-a})}.
\item\label{Prop-c}  A third category of proposals has, as a final state, a single photon which could have its  origin in either  production via coherent nucleus scattering~\cite{Hill:2009ek,Wang:2013wva}, or an anomalous magnetic moment~\cite{0902.3802,1009.5536}, or other new physics process occurring in the detector  involving a photon as a decay product~\cite{Magill:2018jla,Fischer:2019fbw, Datta:2020auq,2110.11944}. These are also based on the assumptions~\ref{C-b-i})$-$\ref{C-b-iv}) as in category~{\bf\ref{Prop-b})} above.
\end{enumerate}
All of the proposals listed above are subject to strong and multiple constraints,  which is expected given the low energies involved, the solidity of the SM and the multiplicity of experiments which have tested it over five decades. Indeed, it has been shown that many of the proposals are disfavoured or in significant  tension with data. To embark on a discussion of  all possible bounds on them and their ramifications  is beyond the scope of our paper and of our aim in this section, but we briefly mention some important classes of constraints below along with some relevant references:
\begin{itemize}
\item With respect to the sterile-active oscillation proposals in category~{\bf\ref{Prop-a})} above, it has become evident over time that sterile neutrinos which can account for appearance data  in short baseline experiments like LSND and MB are in very significant tension with $\nu_\mu$ disappearance data. Additionally, such  neutrinos,  which only interact due to mixing with SM neutrinos,  must contend with stiff constraints from cosmological data~\cite{1905.03254, 2002.07762, 2003.02289}. For a  discussion and complete references for this very active field,   the reader is referred to the review articles~\cite{Dasgupta:2021ies,Dentler:2018sju, 1906.01739, 2109.12385, 1911.03463, 1703.00860}.
\item The other sub-category of solutions in category~{\bf\ref{Prop-a})} which lead to additional $\nu_e$ in the beam due to heavy neutrino decay, introduce new  interacting states. In many cases, such states  affect and can be tested against cosmological observations, such as primordial nucleosynthesis yields and the  total amount of energy density carried by neutrinos at late times. They can also be constrained by considerations of  the supernovae neutrino spectra and the existing bounds on the  flux of antineutrinos from the sun, and from event rates measured at near detectors of many accelerator-based neutrino experiments. For a discussion  and references, see~\cite{Brdar:2020tle,2008.11851}.
\item Proposals in category~{\bf\ref{Prop-b})} are subject to multiple constraints, from near detectors in neutrino experiments, meson decay data, high and ultra-high energy neutrino experiments, colliders, active-sterile mixings, beam dump and dark photon searches, among others.  Discussions and references may be found in~\cite{0901.3589,1011.3046,1110.1610,1502.00477,1511.00683,Coloma:2017ppo,Magill:2018jla,MiniBooNEDM:2018cxm,1810.07185,Arguelles:2018mtc,1904.06787,Coloma:2019qqj,1909.11198,Brdar:2020tle,Abdallah:2020vgg}.
\item Proposed solutions to the anomalies which belong to category~{\bf\ref{Prop-c})}, $\ie$ where a new physics interaction produces a photon in the final state  share many of the same constraints related to mixings between active and heavy sterile neutrinos~\cite{0901.3589,1011.3046,1110.1610,1502.00477, 1511.00683,Magill:2018jla,1904.06787,1909.11198}, since the photon production in many cases is via radiative decay of a heavy neutral lepton. In addition, anomalous magnetic moment constraints~\cite{Corsico:2014mpa,Magill:2018jla,Diaz:2019kim,Studenikin:2021fmn} may be applicable. In some  cases the produced photon is excessively forward, hence is constrained by the observed angular distribution at MB~\cite{Radionov:2013mca}.
\end{itemize}
The new physics proposals in all three aforementioned categories can be tested rigorously in the near future by the Fermilab Short Baseline Neutrino (SBN) program, comprising of three  Liquid Argon Time Projection Chambers (LArTPCs)~\cite{1503.01520} set along the booster neutrino beam at different baselines. These will allow definitive tests of the sterile-active hypothesis (discussed in category~{\bf\ref{Prop-a})} above). In addition, their excellent capability to collect ionization electrons from charged particles passing through the argon in the LArTPC  will enable  detailed track reconstruction, measurements of deposited energy and particle identification for all three types of final states characterizing the new physics proposals.

Overall, at this point in time, it is reasonable to state that the case for new physics arising from an understanding of the LSND, MB and now MicroBooNE results  is strong, as opposed to the case for the results being due to SM backgrounds or systematic errors.  While SM explanations of the excesses are possible, we note that since the first reports of an excess in LSND about 25 years ago,  almost every new  result from these  experiments, whether it has been a measurement or an estimate  of an important SM background or the gradual accumulation of evidence for the excess observed, has strengthened, not weakened the case for new physics. Such  physics may lie among the many proposals covered by categories~{\bf\ref{Prop-a})},~{\bf\ref{Prop-b})}~and~{\bf\ref{Prop-c})} above, or could be  some as yet unimagined new possibility.  

 If, indeed, the excesses are due to new physics, it is difficult to over-estimate the import and breadth of the consequences that would ensue. We make a brief (and incomplete) attempt to give some idea of their range and impact by listing the physics consequences of some of the proposals listed above: 
 \begin{itemize}
 \item A $4.6\sigma$ discrepancy between early-universe and late-universe measurements of the Hubble parameter $H_0$ is currently at the forefront of cosmology~\cite{2001.03624}.  Light sterile neutrinos or, more generally, a hot dark matter component, can  alter the expansion rate at recombination and hence affect the calibration of  the standard ruler with which such measurements  infer distances. They can, thus, potentially reconcile the discrepancy between these measurements~\cite{1307.7715,1308.3255,1906.10136}. The mass and mixing ranges  integral to the simple vacuum oscillation proposals in category~{\bf\ref{Prop-a})}, however, are found to be in strong  tension~\cite{1905.03254,2002.07762,2003.02289} with those required for this reconciliation. A possible resolution entails the existence of new ``secret" interactions between neutrinos~\cite{1310.5926,1310.6337}. 
 \item Sterile neutrinos and/or  singlet scalars, which are components of most proposals in categories~{\bf\ref{Prop-a})},~{\bf\ref{Prop-b})}~and~{\bf\ref{Prop-c})}  both naturally  inhabit the dark sector, which may constitute the bulk of matter in the universe. An understanding of the anomalies may then open a portal to the hugely important but unresolved question about the nature of dark matter. 
 \item In general, sterile neutrinos in the mass ranges relevant for category~{\bf\ref{Prop-a})}  above  impact  supernovae neutrino spectra~\cite{Kolb:1987qy,Shalgar:2019rqe} and high energy astrophysical neutrino emission in substantial ways~\cite{1909.05341,2002.10125,2007.07866}.
 \item Almost all proposals that fall in category~{\bf\ref{Prop-b})} above introduce new interactions beyond the SM mediated by a new vector or scalars, resulting  in non-trivial consequences for future measurements in collider as well as non-collider experiments.  The specifics of each, of course,  depend upon the model proposed. 
 \item Many proposals impact  other important unresolved issues like the  electron~\cite{Morel:2020dww} and muon $g-2$~\cite{Bennett:2006fi, 2104.03281,2104.03247} anomalies, the origin of neutrino masses,  or the KOTO observations~\cite{Shiomi:2021oht}.
\end{itemize} 
\section{The  up-scattering interaction,  cross section and event rate}
\label{sec3}
\begin{figure*}[h!]
\begin{center}
\begin{tabular}{ccc}
\vspace*{-1cm}
\hspace*{-0.6cm}\includegraphics*[width=0.35\textwidth]{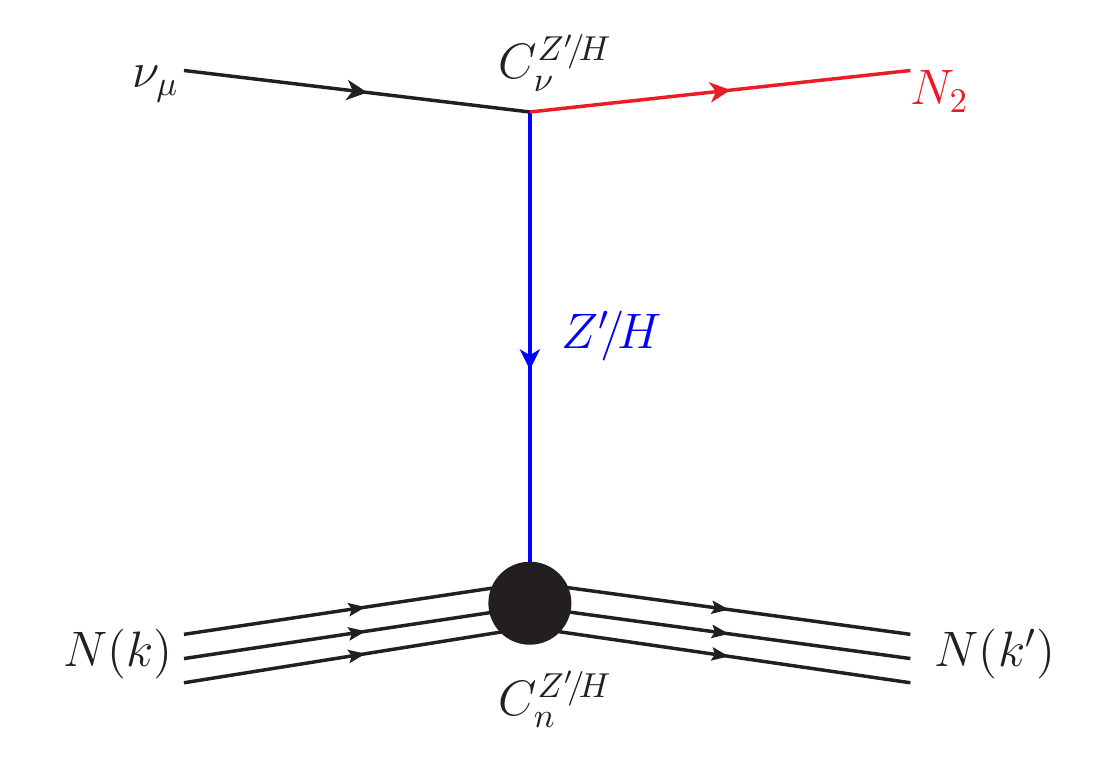} &
\includegraphics*[width=0.35\textwidth]{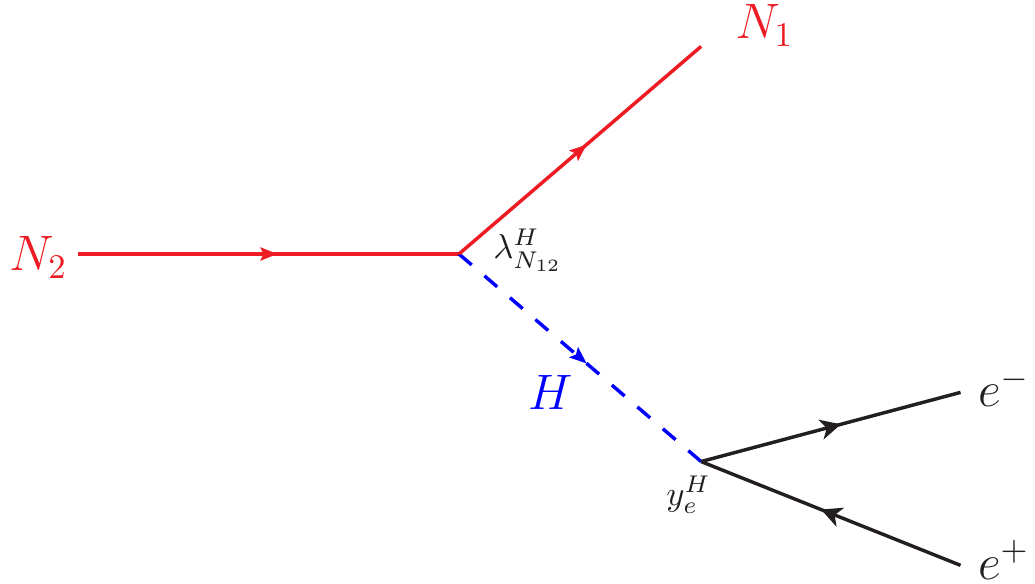} & 
\hspace*{-0.2cm}\includegraphics*[width=0.3\textwidth]{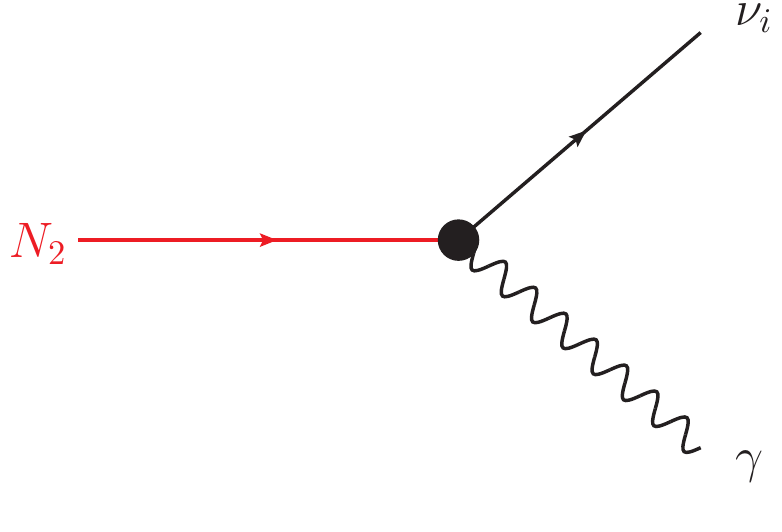} \\[1.2cm] 
 {\footnotesize (a)} & {\footnotesize (b)} & {\footnotesize (c)}
\end{tabular} 
\end{center}
\vspace*{-0.3cm}
\caption{Feynman diagrams for the production and the subsequent decay of the $N_2$  in LSND and MB.} 
\label{FD} 
\end{figure*}

Focusing on categories~{\bf\ref{Prop-b})}~and~{\bf\ref{Prop-c})}, for our examination, we break up the interaction into two parts. We first  consider the tree-level process leading to the up-scattering of an incoming muon neutrino, $\numu$, to a  heavy neutral lepton ($N_2$) in the neutrino detector as shown in figure~\ref{FD}(a), with the underlying assumption that it subsequently decays promptly in the detector. We note that conclusions drawn from this first interaction are applicable to both types of final states, $\ie$ $e^+e^-$ pairs or photons. Final state-specific  considerations related to  the subsequent decay of the $N_2$ (figures~\ref{FD}(b)~and~\ref{FD}(c))  are examined separately in  section~\ref{sec7}.

The mediator for the up-scattering process could be either a light neutral vector boson $Z'$ or a light CP-even scalar $H$. The relevant interaction Lagrangian for the up-scattering process in each case  is given by
\begin{eqnarray}
\mathcal{L}^{Z'}_{\rm{int}} \!\!&=&\!\! (C^{Z'}_{\nu} \bar{\nu}_{iL} \gamma^{\mu} N_{Lj}Z'_{\mu} +h.c.) +  C^{Z'}_{n} \bar{U}_{n} \gamma^{\mu}U_n Z'_{\mu},\\
\mathcal{L}^H_{\rm{int}} \!\!&=&\!\! (C^H_{\nu} \bar{\nu}_{iL} N_{Rj}H +h.c.) + C^H_{n} \bar{U}_{n} U_n H,
\end{eqnarray}
where $U_{n}$ is the nucleon field and $i, j=1,2,3$.\footnote{We have assumed the presence of three right-handed neutrinos, as is usual for generating neutrino mass,  although our interaction only requires two such particles ($N_1$ and $N_2$). Additionally, these neutrinos are assumed to be Majorana in nature, which is relevant when we consider the decay of $N_2$ in what follows. We note that angular distributions for  the decay of Majorana neutrinos differ from those for Dirac neutrinos. Specifically,  in the Majorana case,  the angular distribution of the daughters in the decay of a heavy neutrino into a lighter neutrino  and a self-conjugate boson is isotropic in the parent’s rest frame~\cite{Balantekin:2018ukw, BahaBalantekin:2018ppj,Kayser:2018mot}, a fact that we use in our considerations.}  For simplicity, we assume that whether the mediator is a  $Z'$ or  an $H$, its coupling to a  proton is  the same as its  coupling to a neutron. The up-scattering cross section depends on the overall product of the coupling constants $C^{Z'}_{\nu} C^{Z'}_n$ ($C^{H}_{\nu} C^H_n$) for the vector (scalar) mediator. 

\begin{figure*}[t!]
\begin{center}
\includegraphics[width=0.5\textwidth]{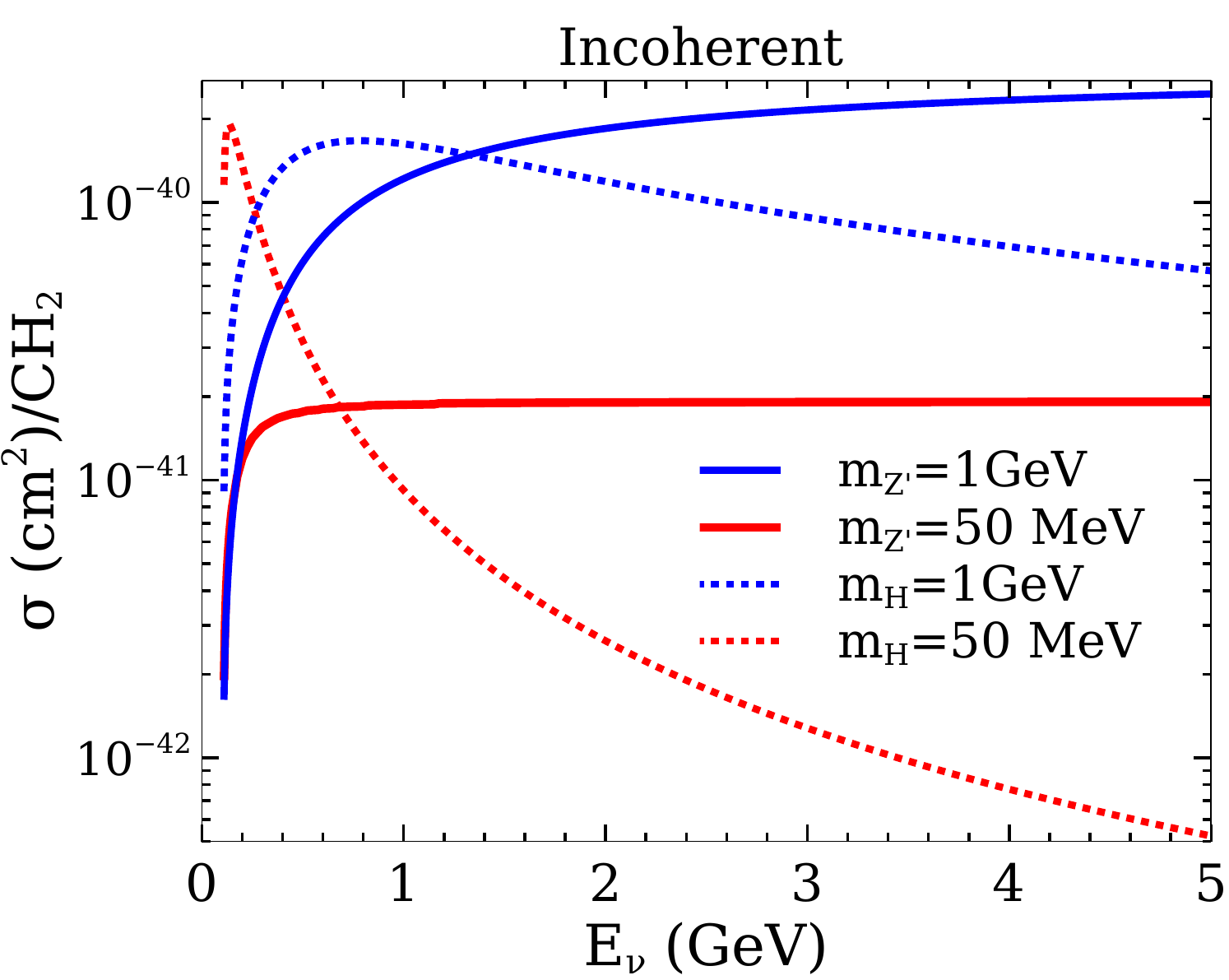}~~
\includegraphics[width=0.5\textwidth]{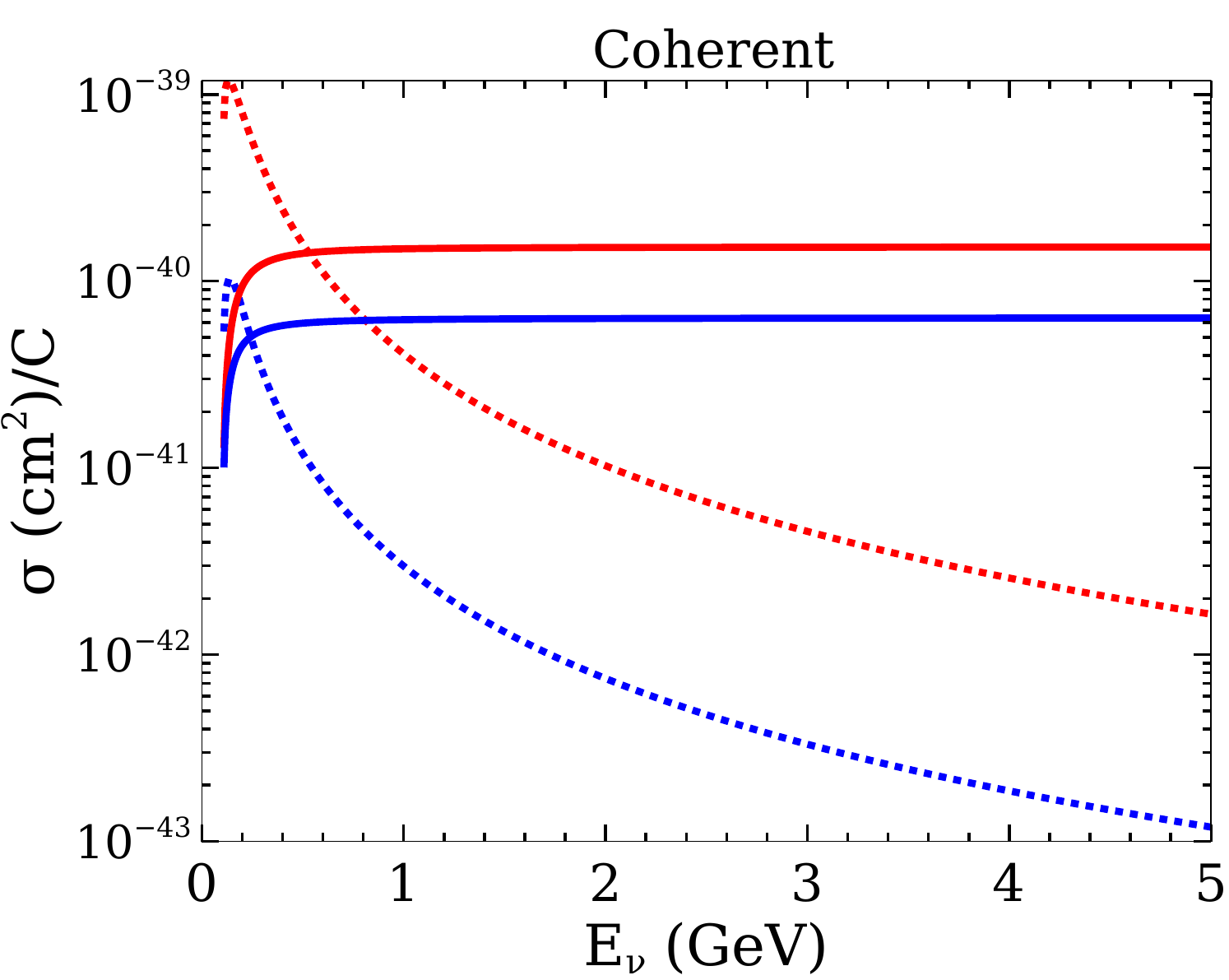}
\caption{The incoherent (coherent) cross section per CH$_2$ molecule (C atom) as a function of incoming neutrino energy. The overall constants for different kinds of mediator masses are taken from Table~\ref{tab}.} 
\label{x1}
\end{center}
\end{figure*}
The total differential cross section for the production of $N_2$ mediated by a vector $Z'$ or a scalar $H$, for the target in MB, $\ie$, CH$_2$, is given by

\begin{equation}
\Big[\dfrac{d\sigma^{Z'\!/\!H}}{dE_{N_2}}\Big]_{{\rm CH_{2}}}=\Big[\underbrace{14(C^{Z'\!/\!H}_{\nu} C^{Z'\!/\!H}_{n})}_{\textrm{\footnotesize{{incoherent}}}} 
+\underbrace{ 144(C^{Z'\!/\!H}_{\nu} C^{Z'\!\!/\!H}_{n})^{2}  e^{-2b|q^{2}|}}_{\textrm{\footnotesize{{coherent}}}}\Big]\dfrac{d\sigma^{Z'\!/\!H}}{dE_{N_2}}.
\label{tot_xsec}
\end{equation}
The coherent part of the cross section receives contributions from the entire  carbon nucleus (C$^{12}$),  and these decrease  as $|q^2| = |(k'-k)^2| $ increases.  
We implement this by using  the form factor $exp(-2b|q^{2}|)$~\cite{Hill:2009ek}, where  $b=25$~GeV$^{-2}$~\cite{Freedman:1973yd, Hill:2009ek}. The number of $N_2$ in the final state is given by
\begin{equation}
{\rm N}^{Z'\!/\!H}_{N_2} =\eta\,\int dE_{\nu}\, dE_{N_2}\dfrac{d\Phi^{\rm{\nu}}}{dE_{\nu}} \, \Big[\dfrac{d\sigma^{Z'\!/\!H}}{dE_{N_2}}\Big]_{{\rm CH_{2}}},
\label{N2events}
\end{equation}
where $\Phi^{\rm{\nu}}$ is the incoming muon neutrino flux, $E_{N_2}$ and $E_{\nu}$ are the respective energies and $\eta$ is the {signal} efficiency of the detector. 

A substantial amount of information about the viability of various solutions can be gleaned from focussing simply on the production of $N_2$ in the final state. To calculate the total number of $N_2$, we use the  signal efficiencies of the LSND and MB experiments provided in~\cite{LSND:2001aii, MiniBooNE:2020pnu} and references therein. The actual signal in the detector depends on the mechanism of the decay of $N_2$ to a $\gamma$ or  an $e^+ e^-$ pair. However, for our purpose in this section, we assume that the signal arises from  $N_2$ decay in the detector, and adjust the values of the couplings so as to produce a total of $560$ $N_2$ in the final state. This number corresponds to the central value  of the excess observed~\cite{MiniBooNE:2020pnu}   over the SM expectation during the neutrino run, given the total exposure so far,  and represents, in some sense, the minimum number of $N_2$ that should be produced. The events are calculated  using eqs.~(\ref{tot_xsec}),(\ref{N2events}) and the assumed values of the product of couplings  are provided in Table~\ref{tab}.

\begin{table}[t!]
\begin{center}	
 \begin{tabular}{|c|c|c|} 
 \hline\Tstrut
 Mediator Mass & $C^{Z'}_{\nu} C^{Z'}_n$  &$C^{H}_{\nu} C^{H}_n$  \\   
 \hline\Tstrut
 50~MeV & 1.04$\times 10^{-8}$ & 9.3$\times 10^{-8}$ \\ 
 \hline\Tstrut
 1 GeV  & 8.5$\times 10^{-7}$ & 2.14$\times 10^{-6}$ \\
 \hline\Tstrut
  300~MeV & $-$ & 5.35$\times 10^{-7}$ \\
 \hline
\end{tabular}
\caption{The values of the overall couplings for the vector and scalar mediators for different values of mediator masses to produce 560 $N_2$ in the MB final state. The mass of $N_2$ is 100~MeV.}
\label{tab}
\end{center}
\end{table}
Finally, it is appropriate to comment on  the role that nuclear effects may play in scenarios of new physics that involve incoherent scattering, like the ones considered here. For known neutrino-nuclear cross sections, several techniques are in use~\cite{Gaisser:1986bv,Kim:1994zea,Kim:1996az,Singh:1993rg,Engel:1993nn,Marteau:1999jp}. At lower energies ($\eg$ at LSND) shell-model techniques yield good results. At somewhat higher, $\ie$ intermediate energies  (several hundred MeVs to a few~GeV) the Continuum Random Phase Approximation (CRPA) has been successfully used to show that nuclear effects are relevant mostly at low momentum transfers, while at higher energies the Fermi  gas model has worked well. Overall, these techniques have been used in conjunction with experimental data in order to obtain a better understanding of these interactions. Additionally, for energies up to $\sim 1$~GeV, the role of final state interactions may be relevant~\cite{Bleve:2000hc,Co:2002jiq}. While it is reasonable to assume, as we do here,  that similar considerations will apply in the case of new physics scenarios, only further studies can confirm the validity of such an assumption. We do not undertake them here, as they would be outside the scope of this paper\footnote{We have, nonetheless, put in a cut of a minimal requirement of 8~MeV for a neutron to exit the nucleus.}.

\section{Requirements resulting from the variation of the cross section with incoming neutrino energy and with mediator mass}
\label{sec4}
The variation of the  up-scattering cross section with incoming neutrino energy  depends on the type of mediator, as shown in figure~\ref{x1}. The left panel shows the incoherent cross section while the right panel corresponds to the coherent cross section for vector ($Z'$, solid curves) and scalar ($H$, dotted curves) mediators for two values of the mediator mass, (50~MeV (red) and 1~GeV (blue)) for a CH$_2$ target. 

We see that the scalar and vector mediated cross sections behave distinctly, and our representative calculations bring out the following qualitative points:
\begin{itemize}
 \item While in all cases the cross section initially rises as the energy is increased from its lowest values,  it subsequently drops for a scalar mediator whereas it remains approximately flat with increasing energy for a vector mediator. Both the coherent and incoherent parts exhibit this general behaviour. It is this relatively rapid drop in the cross section  for $H$ as the neutrino energy rises that allows solutions with a scalar mediator to comfortably
 avoid constraints coming from CHARM~II~\cite{CHARM-II:1994dzw} and  MINER$\nu$A \cite{Valencia:2019mkf},  compared to the $Z'$ mediated process. We recall that the $\nu-e$ scattering measurements in both these experiments can  in principle put tight constraints~\cite{Arguelles:2018mtc} on the model that explains the LEE. The coherent signal seen by them,  ($\nu_{\mu} + A \rightarrow N_2 + A \rightarrow N_1 +e^+ + e^- + A) $ could mimic a charged or neutral current $\nu-e$  event. Hence, a large  coherent contribution, as is present in the $Z'$ case,  will be in conflict with the $\nu-e$ scattering measurements of CHARM~II and MINER$\nu$A. 
 \item We also note  from figure~\ref{x1} that the coherent contribution dominates over the incoherent part for lighter mediator masses, whereas the opposite is true for the higher mass choice for both types of mediators. Thus lighter mediators tend to make large coherent contributions, and since these tend to be forward in angle, they help populate event bins for $\cos \theta\simeq 1$, a point that we explore further in section~\ref{sec5}.
\item We note that for  LSND, contributions to events stem from the incoherent part of the cross section only, given the presence of a neutron in the final state. We thus focus on  the region in the left panel of  figure~\ref{x1}, and note the behaviour as  the  energy drops from MB ($\sim 800$~MeV) to LSND DAR flux values ($\sim 150-200$~MeV). We see that for the higher mass mediators ($m_{Z'\!/\!H}=1$~GeV; blue curves), while the incoherent cross section drops  for both mediators,  the vector cross section has lower values to begin with compared to the scalar and also drops rapidly.  For instance, it can be seen that the cross section for the $Z'$ drops an order of magnitude over this energy range for $m_{Z'}= 1$~GeV.

 For the lighter mass choices ($m_{Z'\!/\!H}=50$~MeV; red curves), the incoherent scalar cross section is significantly higher than the vector one over this energy range, and in fact increases as the energy is lowered, unlike its vector counterpart. This reduction  in the incoherent vector cross section at  values below MB energies ($< 800$~MeV) makes it more difficult for models with an additional vector mediator to give a sufficient number of electron-like excess events at LSND, even though a high enough $Z'$ mass may allow one to successfully evade the CHARM~II and  MINER$\nu$A bounds.  On the other hand, too low a scalar mediator mass  results in many more events than those observed in LSND,  
 both in the $20-60$~MeV visible energy range which recorded data, and beyond $60$ MeV, where only a limited number of events were seen.
 \item Finally, we point out an important constraint that applies to models which use scalar mediators, especially those with low masses $m_H\simeq 100$~MeV. As  can be seen in figure~\ref{x1} the cross section tends to rise  at low values of the incoming neutrino energy. However, in such models, if $N_2$ decays primarily invisibly, as in~\cite{Dutta:2020scq}, the incoherent  interaction would mimic the neutral current interaction $\nu N \rightarrow\nu N$, which has been measured at MB~\cite{0909.4617} at these energies, and found to be in agreement with the SM. Conformity with this measurement  is thus an important restriction on such models.
 \end{itemize}
  Overall, the cross section and mediator mass considerations  for a common solution thus appear to favour scalar mediators over vectors.  Secondly, our representative calculations also point to a preference for  lighter (but not ultra-light)  mediators if both excesses are to have a simultaneous solution. 
  
  We next provide some example numbers  to quantitatively illustrate these qualitative conclusions:\\
  
Assuming a mediator mass of $1$~GeV, we  fit the MB events by considering the appropriate values of $C^{Z'}_{\nu} C^{Z'}_n$ and $ C^{H}_{\nu} C^H_n$. Now using the same coupling values, we calculate the number of produced $N_2$ of mass $100$~MeV in LSND.
 We find that:
\begin{itemize}
\item To produce $560$ $N_2$ in final state of MB, the required value of  $C^{Z'}_{\nu} C^{Z'}_n$ is $8.5\times 10^{-7}$. These  couplings, yield  around 7 $N_2$ in LSND instead of  the required number of 32~\cite{LSND:2001aii}.
\item For the scalar mediator, the necessary value  of $C^{H}_{\nu} C^{H}_n$ is $2.14\times 10^{-6}$, obtained from the MB fit. Using the same coupling constant, we get around 35 events in LSND. 
\end{itemize}
We see in this example that while both mediators fit MB total events, the scalar mediator produces 5 times more events compared to the $Z'$ in LSND.

Additionally,  as mentioned above, the different behaviour exhibited by the vector also places more restrictions on it constraint-wise than it does on the scalar. Noting, from the right panel of figure~\ref{x1}, 
that the coherent contribution to the cross section  increases as the mediator mass decreases,  we find that the mass of $Z'$ should  be approximately  500 MeV to  avoid the  CHARM II and MINER$\nu$A constraints discussed above, if it has to concurrently  produce 560 $N_2$ in MB. On the other hand,  no such restriction results for the scalar, since the coherent cross section drops rapidly as it approaches  CHARM~II ($\langle E_\nu\rangle=24$~GeV, $\langle E_{\bar\nu}\rangle=19$~GeV) and MINER$\nu$A  ($\langle E_\nu\rangle=6$~GeV) energies. 

\begin{figure*}[t!]
\begin{center}
\includegraphics[width=0.5\textwidth]{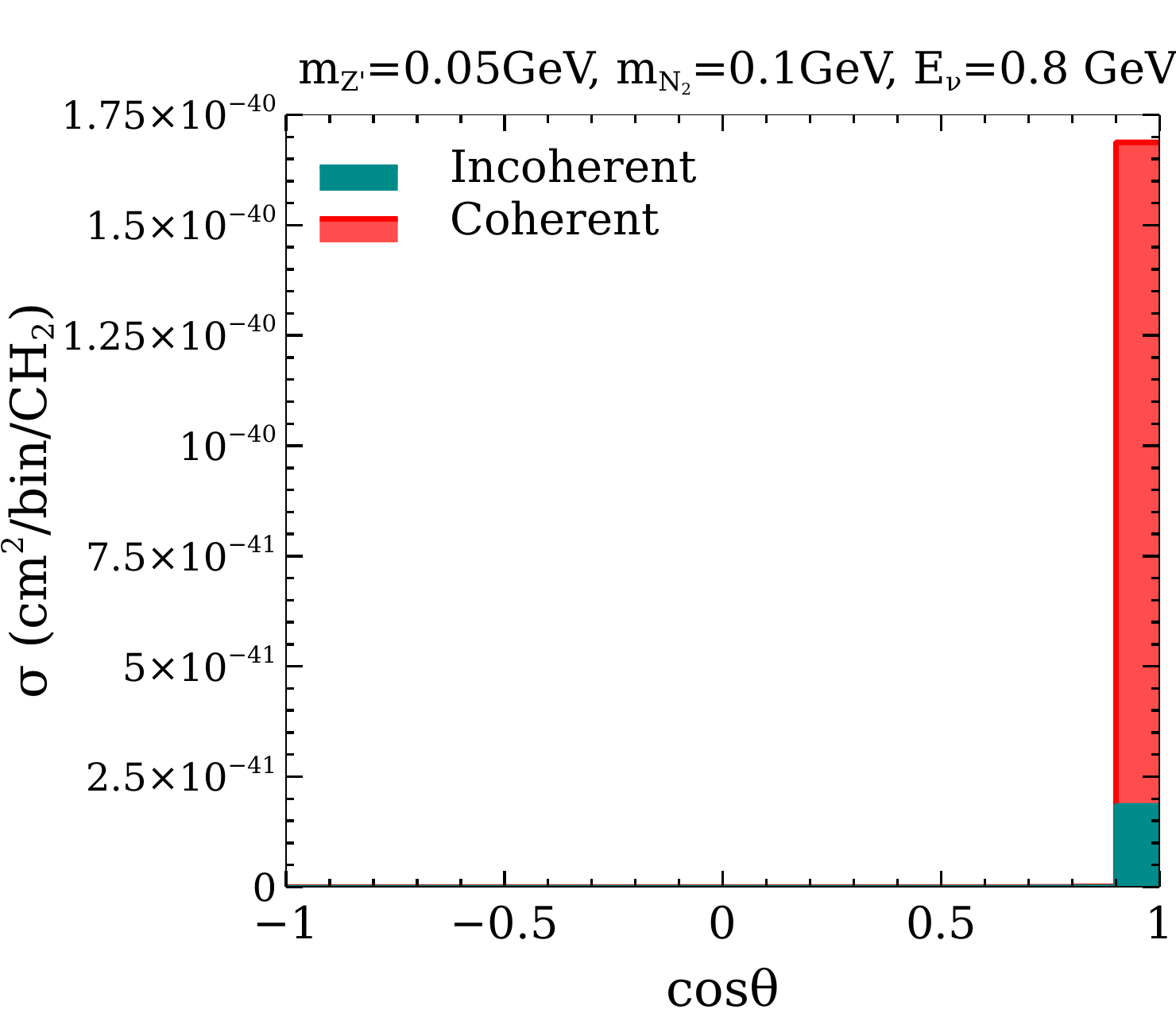}~~\includegraphics[width=0.5\textwidth]{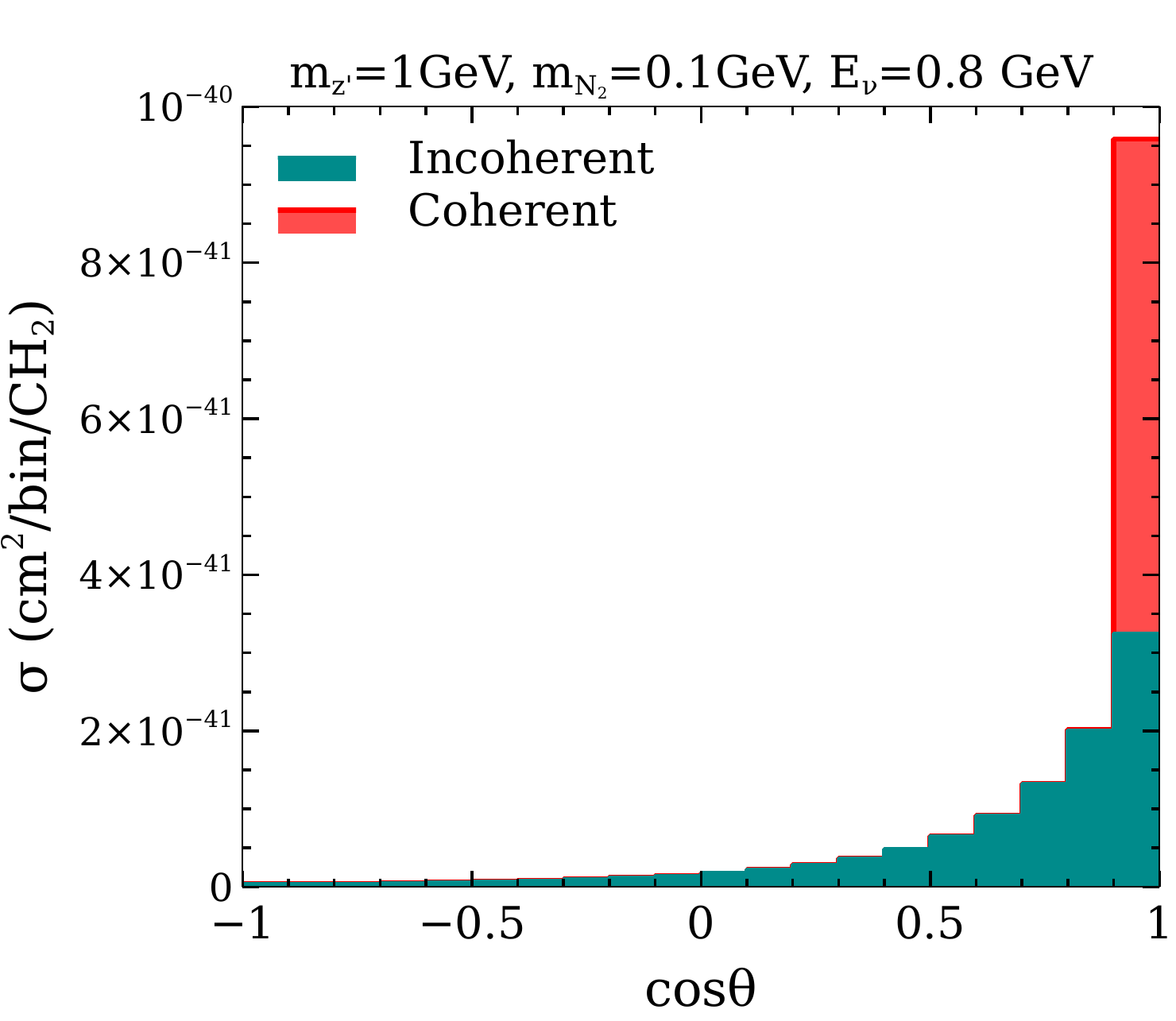}
\includegraphics[width=0.5\textwidth]{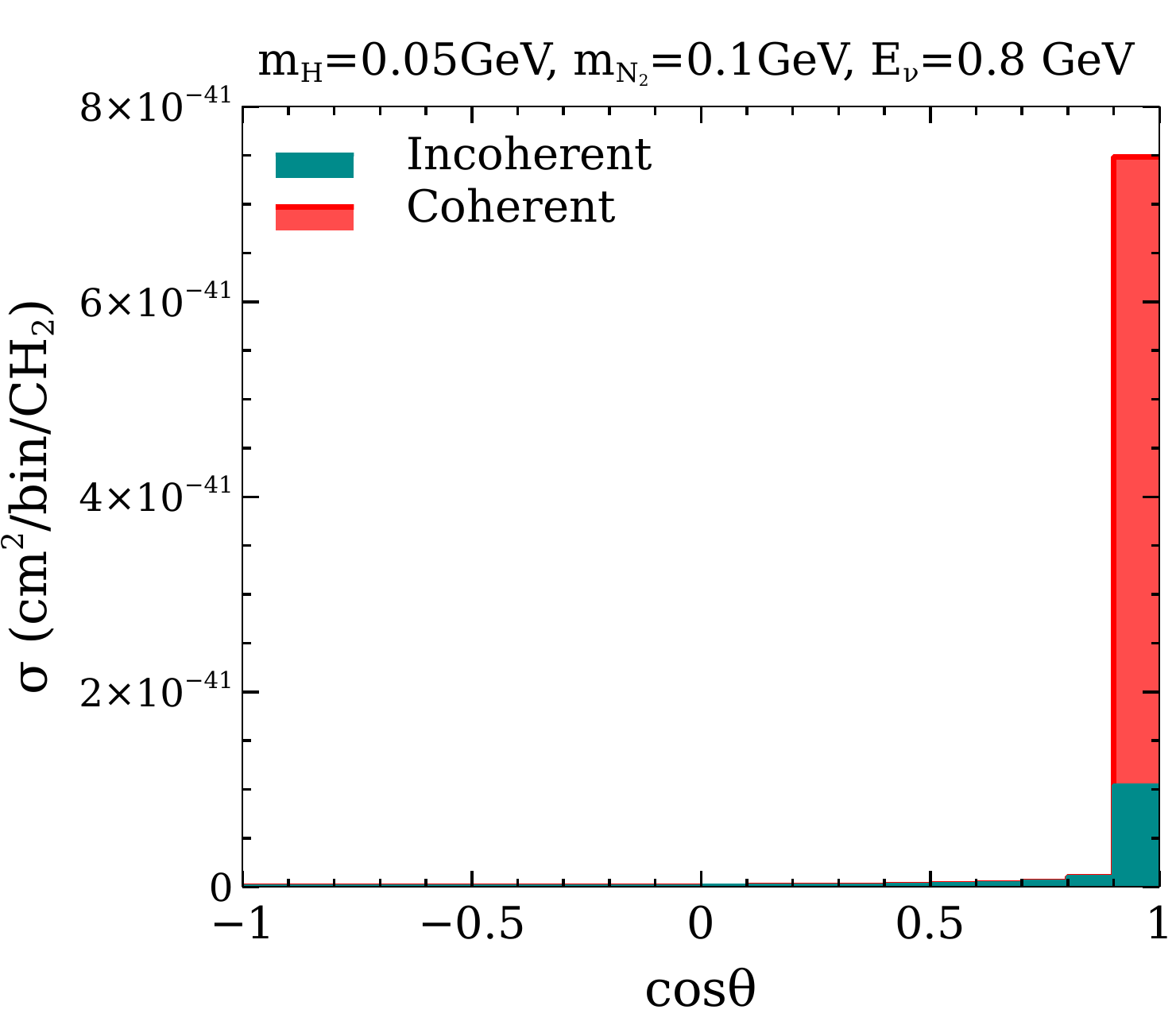}~~\includegraphics[width=0.5\textwidth]{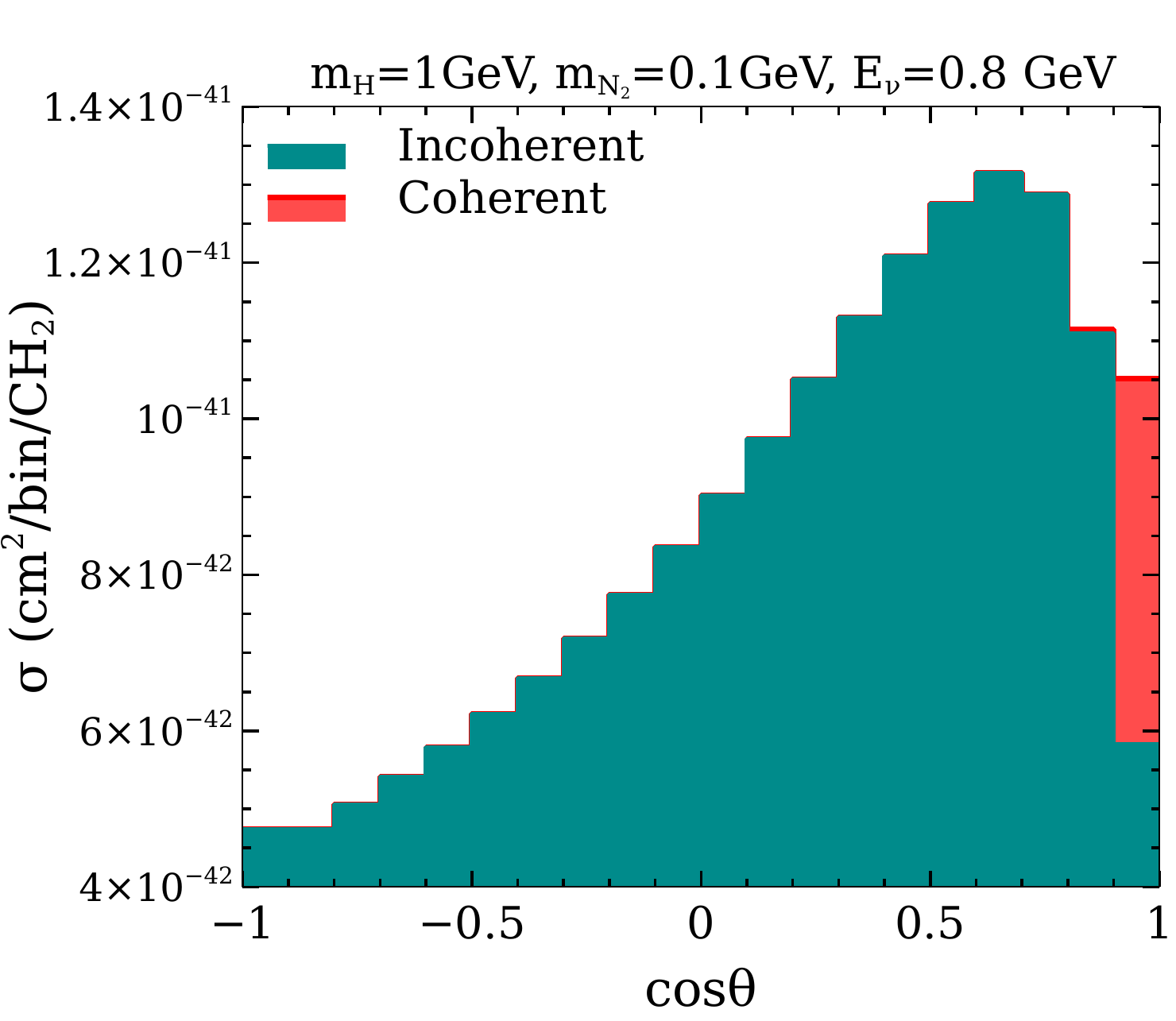}
\caption{Angular variation of the total cross section for two different mass choices of light scalar and light vector mediators. Both coherent and incoherent contributions are shown.}
\label{xsec-angular}
\end{center}
\end{figure*}

\section{Requirements resulting from fitting the angular distribution in MB}
\label{sec5}
An examination of the angular distribution of MB is also useful from the point of view of  imposing requirements on proposed solutions. The  excess in MB  is distributed over all directions, but is  moderately forward.  Figure~\ref{xsec-angular} shows, bin-wise,  the  cross section responsible for the production of $N_2$  as a function of the cosine of the angle between the momentum direction of $N_2$ and the beam direction\footnote{We have checked that this is a good approximate indicator of the eventual angle that the signal will form with the beam, once $N_2$ decays.}. The incoming neutrino energy is fixed at 800~MeV and $m_{N_2}=100$~MeV for all panels.  Both the coherent (red) and incoherent (green) contributions are shown. From the upper and lower left panels, we see that when the mass of the mediator is low, ($50$~MeV), almost all the produced $N_2$ are in the first ($\ie$ most forward) bin for both a scalar and a vector mediators. As the mass of the mediator is increased to $1$~GeV (right panels),  the distribution shifts and the other bins are also populated for both types of mediators, although the shifts are qualitatively different.

 One way to counter balance the effect of a low mass mediator on the angular distribution is to increase the mass of  $N_2$. The energy budget set by the incoming muon neutrino ($E_\nu\sim 1$~GeV) then reduces the kinetic energy of the $N_2$ and drives the distribution towards greater isotropy once it decays to $\eg$ an $e^+ e^-$ pair.  An example of this is provided by the proposed solution to MB  in~\cite{Bertuzzo:2018itn}, where the benchmark point assumes a vector mediator mass of $\sim 30$~MeV and an $N_2$ mass of $\sim 420$ MeV to get the required angular distribution.   A common solution to both the LSND and the MB anomalies, however, is disallowed by such a choice, since the neutrino energies at LSND are much smaller, with no flux beyond $E_\nu\sim 250$~MeV. Thus, while a heavier mediator or a heavier $N_2$ both assist in getting the desired MB angular distribution, a common solution that also helps us understand LSND suggests the  use of an $N_2$ mass closer to $100$~MeV, while using a heavier mediator with a mass $\sim 1$~GeV. Using these trial values in the right panels of  figure~\ref{xsec-angular}, however, shows that a vector mediator  over-produces the forward events ($\ie$ those in the first bin), whereas  the scalar mediator under-produces them.

\begin{figure}[t!]
\begin{center}  
\includegraphics[width=0.5\textwidth]{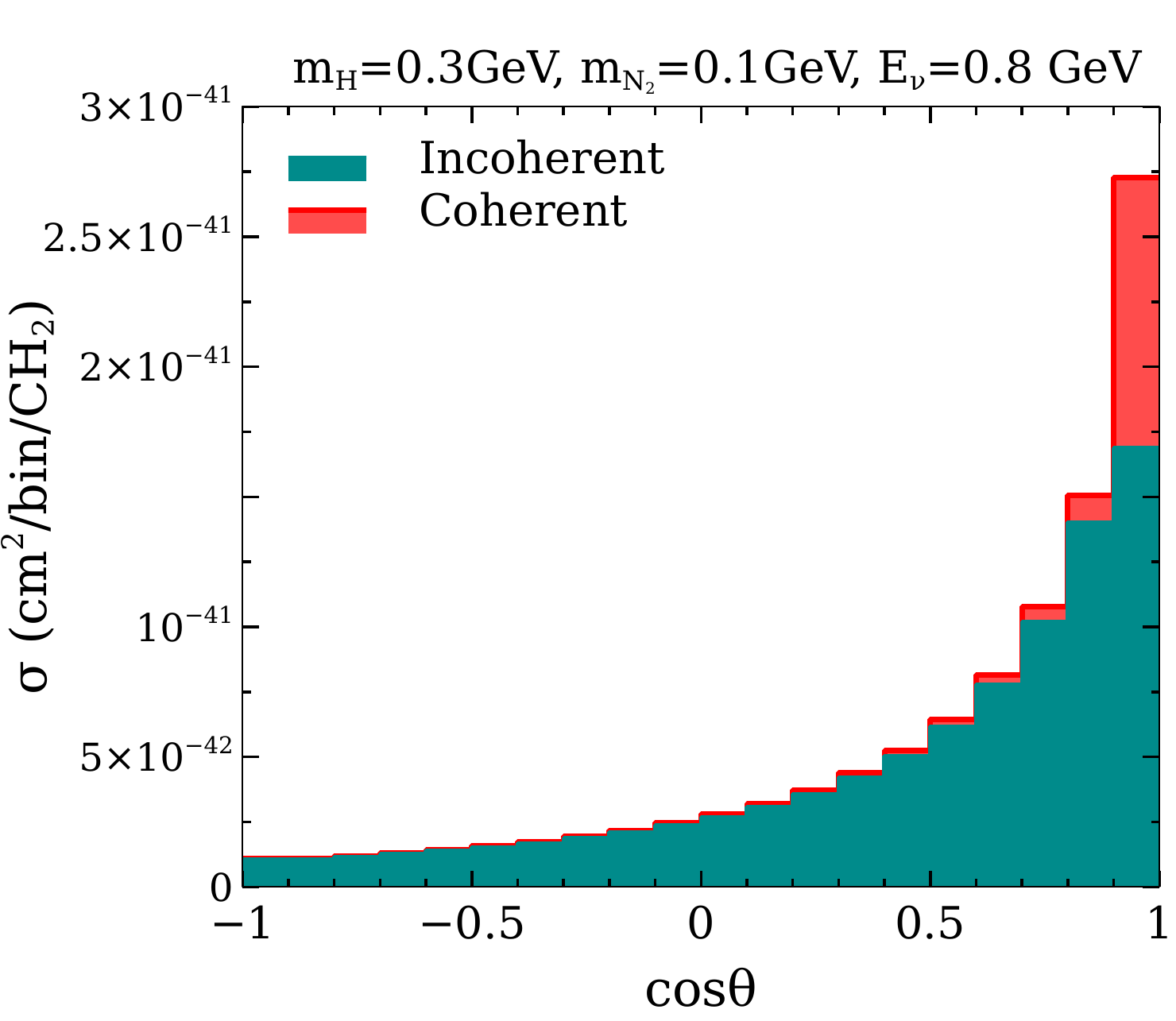}
\caption{Angular variation of the total cross section with an intermediate mass scalar.  Both coherent and incoherent contributions are shown.}
\label{angular-scalar}
\end{center}
\end{figure}

 From the previous section which considers the effects of mediator mass and energy on the cross section keeping constraints in mind, we found that  these criteria point to the scalar mediator as being preferred over the vector. The considerations in this section suggest, among other things, that using a single scalar mediator and adjusting its mass as well as the mass of $N_2$ will allow us to find both a common solution to the two anomalies as well as match the angular distribution in MB. While an $N_2$ mass of 100~MeV is a good ball-park value, one must also keep in mind that as this is lowered, the event rate at LSND rises quickly beyond what was observed. A sample choice that has the potential to achieve an appropriate balance is shown in figure~\ref{angular-scalar}, along with the chosen parameter values; in particular, the scalar mass used is 300~MeV.
  
 Our generic arguments so far have led us to the conclusion that one can achieve a fit to both LSND and MB data using a heavy neutrino of about $\sim 100$~MeV which decays to give the visible signal from an interaction that is mediated by a scalar of mass $\sim 300$~MeV. In general, such a  solution usually  requires the addition of a singlet scalar that mixes with the SM  Higgs boson to generate interactions with quarks and leptons. Constraints on such models arise from a variety of different experiments, and are unusually tight, often forcing the mixing to be very low and making it difficult to explain significant excesses above SM backgrounds  without further additions to the particle spectrum, as, $\eg$ in~\cite{Datta:2020auq}. Some of the important constraints are from NA62~\cite{NA62:2020xlg}, E949~\cite{BNL-E949:2009dza}, CHARM~\cite{Winkler:2018qyg}, NA48/2~\cite{NA482:2016sfh}, LHCb~\cite{LHCb:2016awg,Aaij:2015tna} and Belle~\cite{Belle:2009zue}.
 
Finally, we remark that we have examined the fit to the LSND angular distribution for the 300~MeV scalar considered above. We find that acceptable fits may be possible, but given the uncertainty and limited range of the  measured LSND distributions, it is difficult to draw any firm conclusions based on this, especially as we consider the $N_2$ distributions as an approximation to the actual distribution of the final state.
 \section{Requirements resulting from considerations related to the energy distributions in LSND and MB}\label{sec6} 
\begin{figure*}[t!]
 \begin{center}
\includegraphics[width=0.5\textwidth]{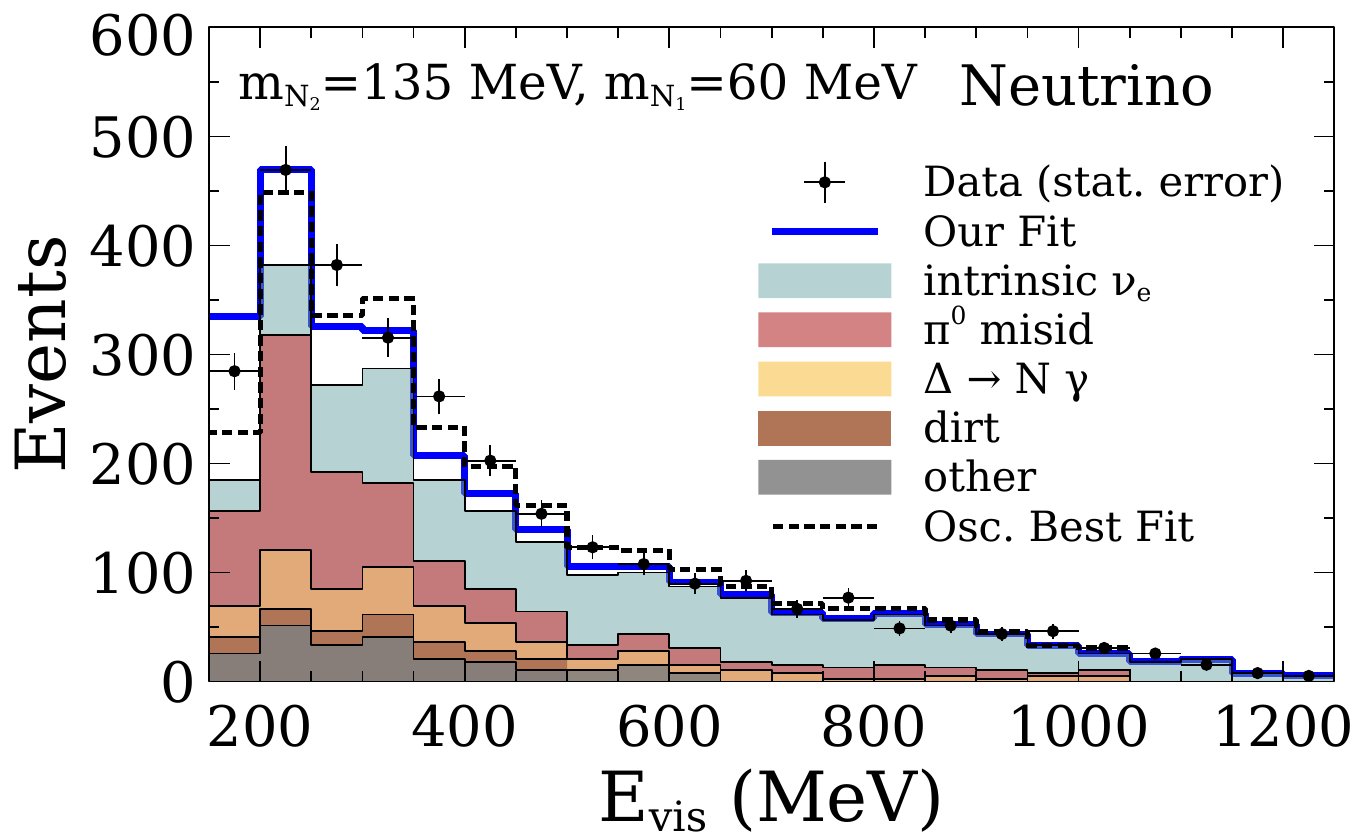}~~
\includegraphics[width=0.5\textwidth]{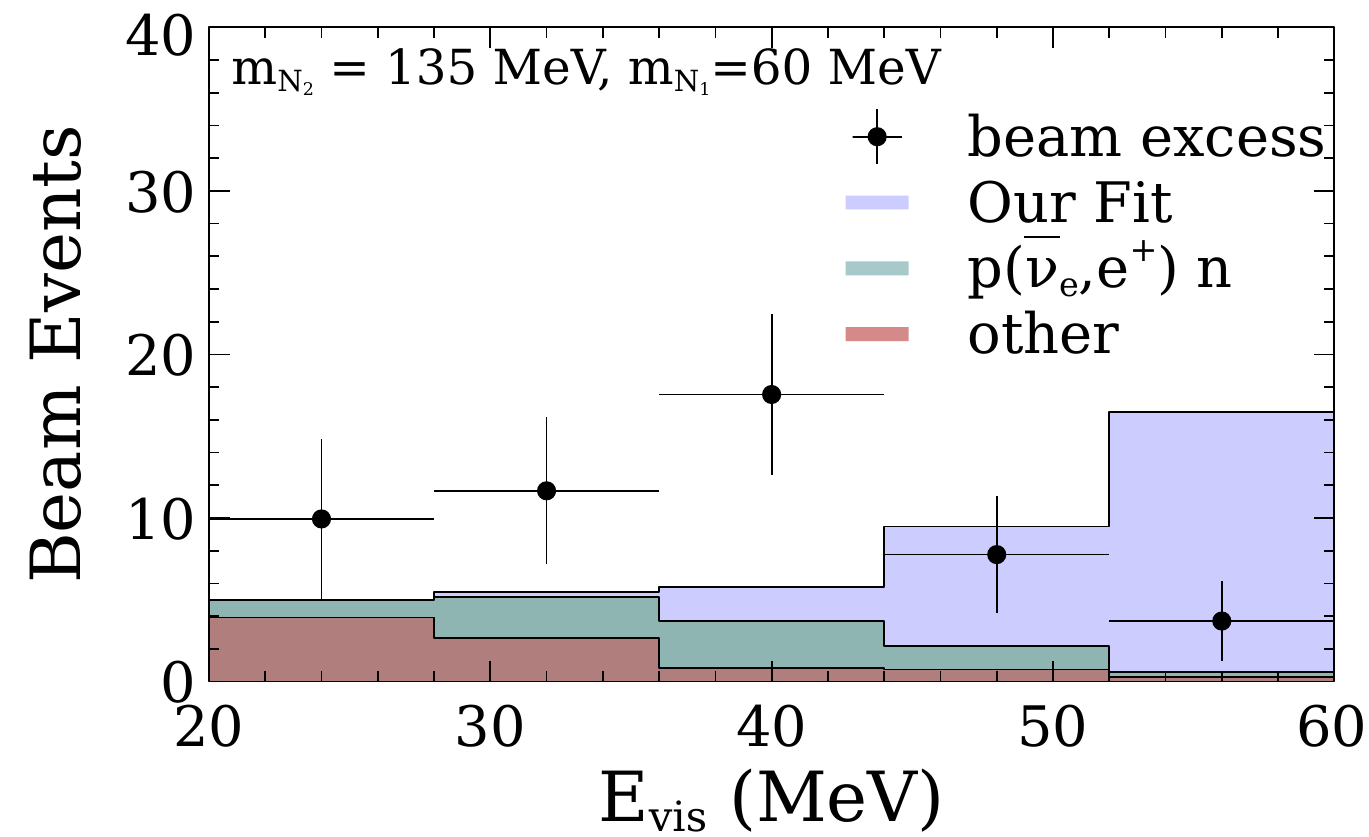}\\[0.5cm]
\includegraphics[width=0.5\textwidth]{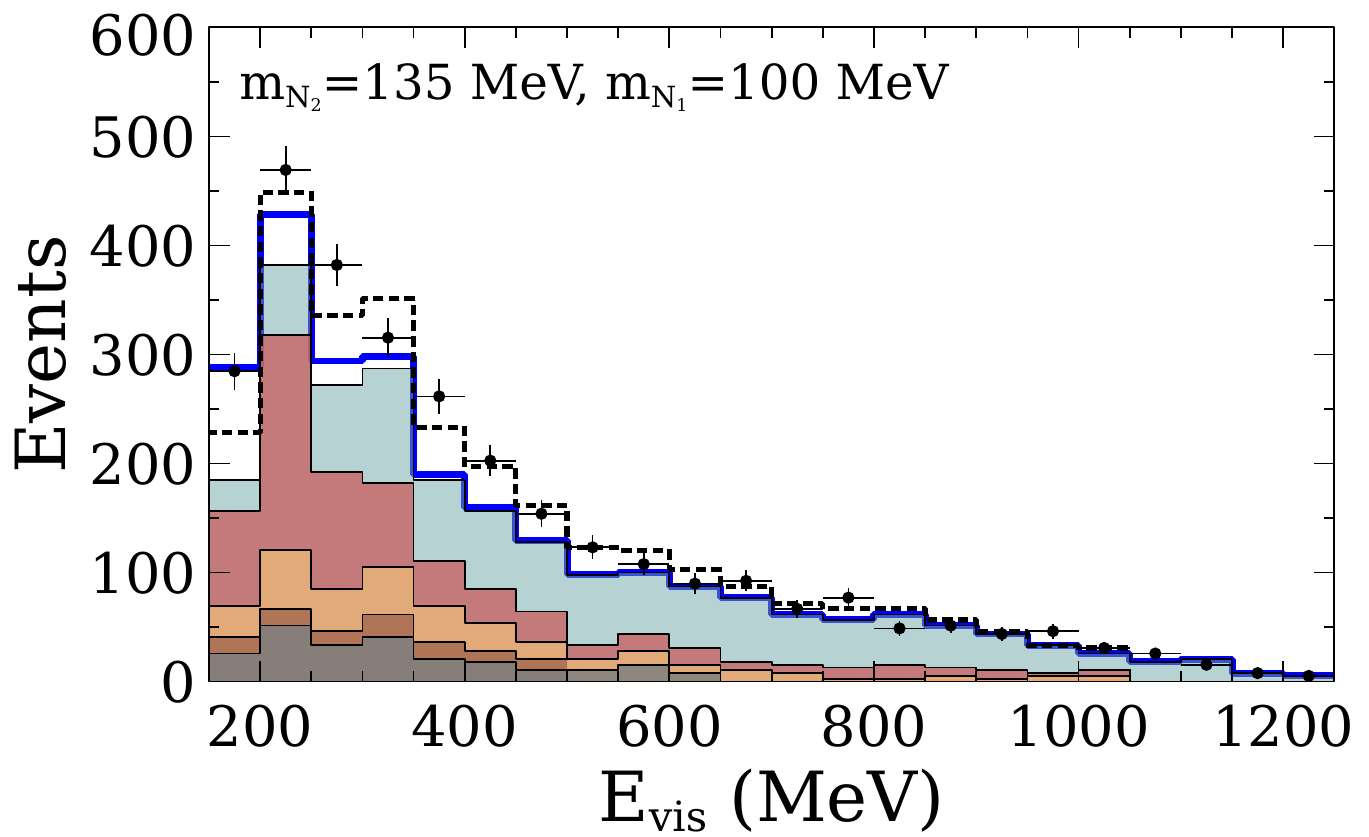}~~
\includegraphics[width=0.5\textwidth]{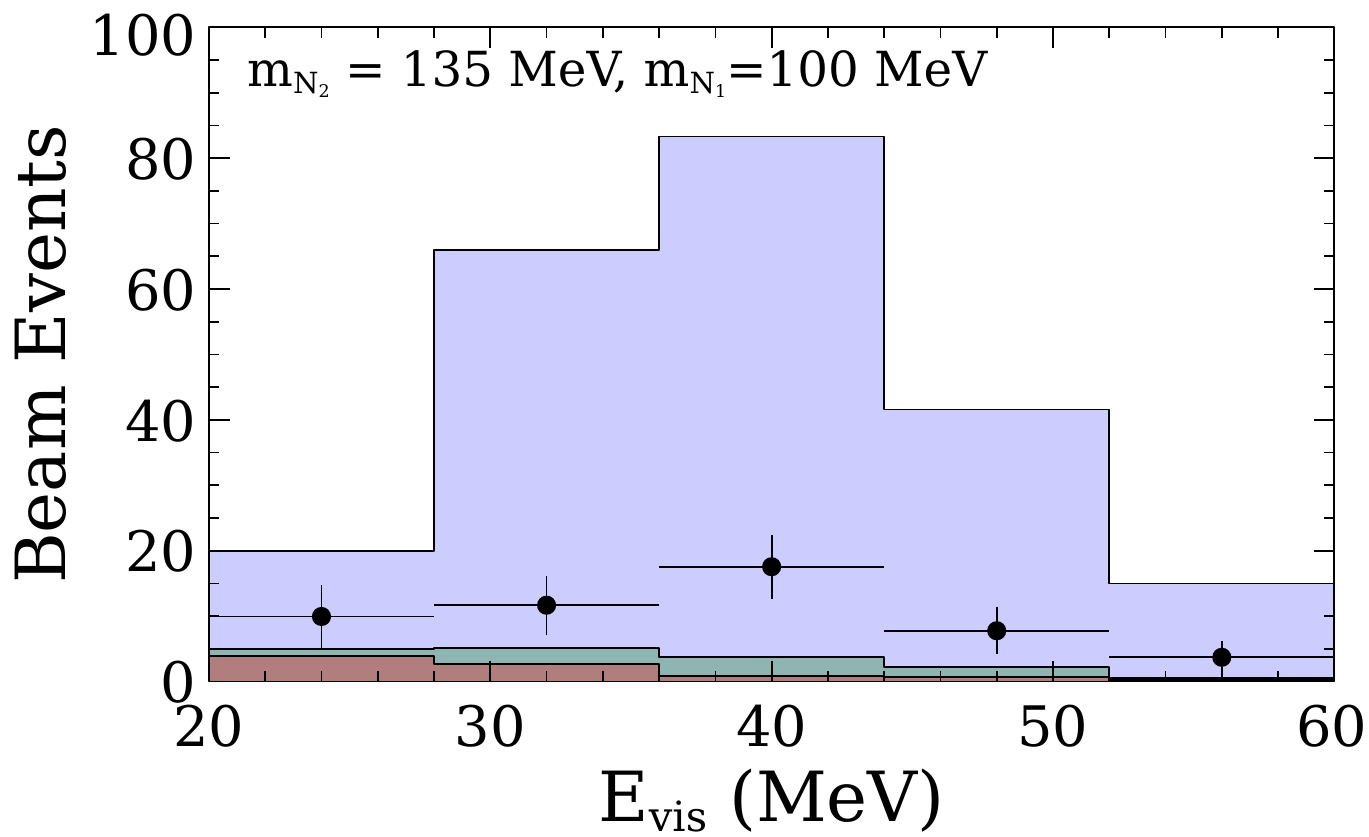}\\[0.5cm]
\includegraphics[width=0.5\textwidth]{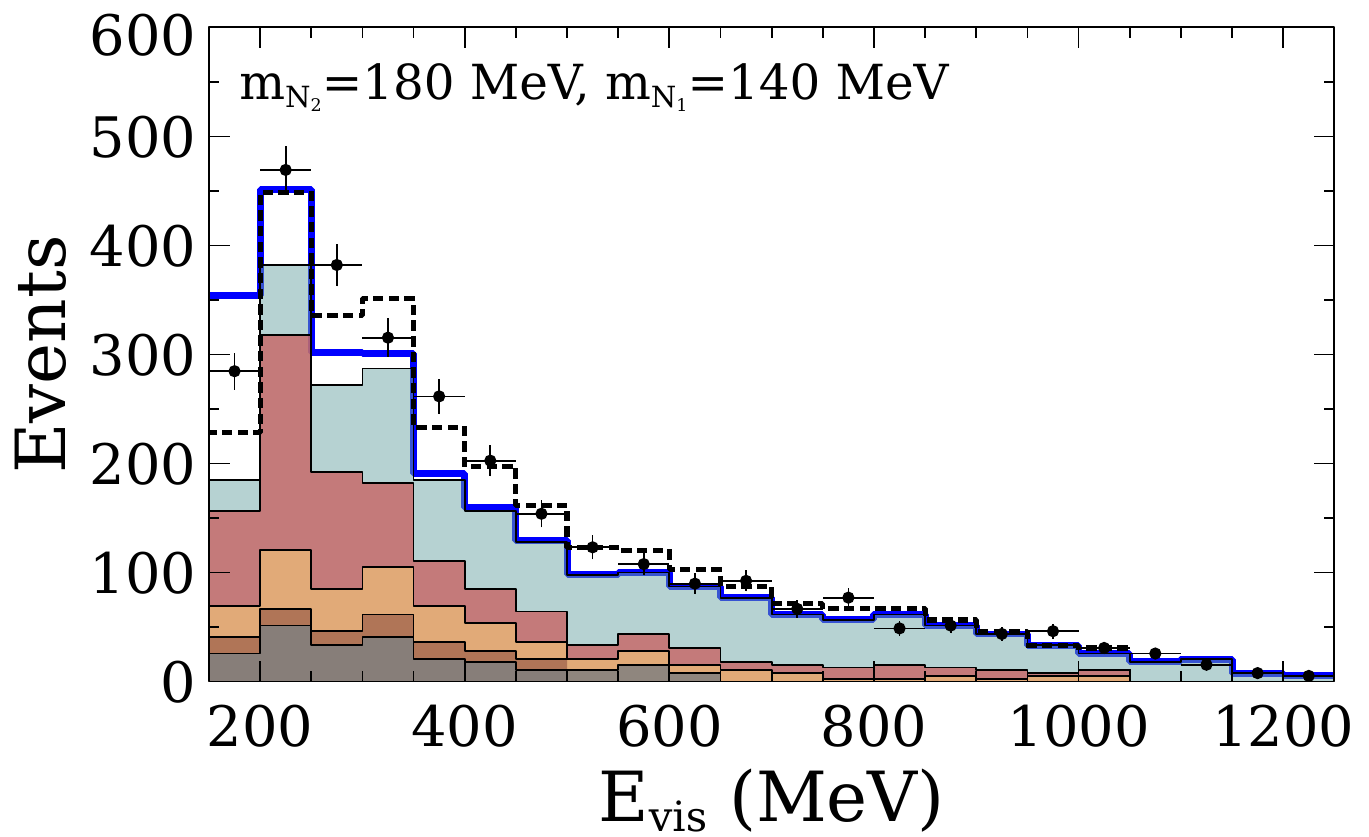}~~
\includegraphics[width=0.5\textwidth]{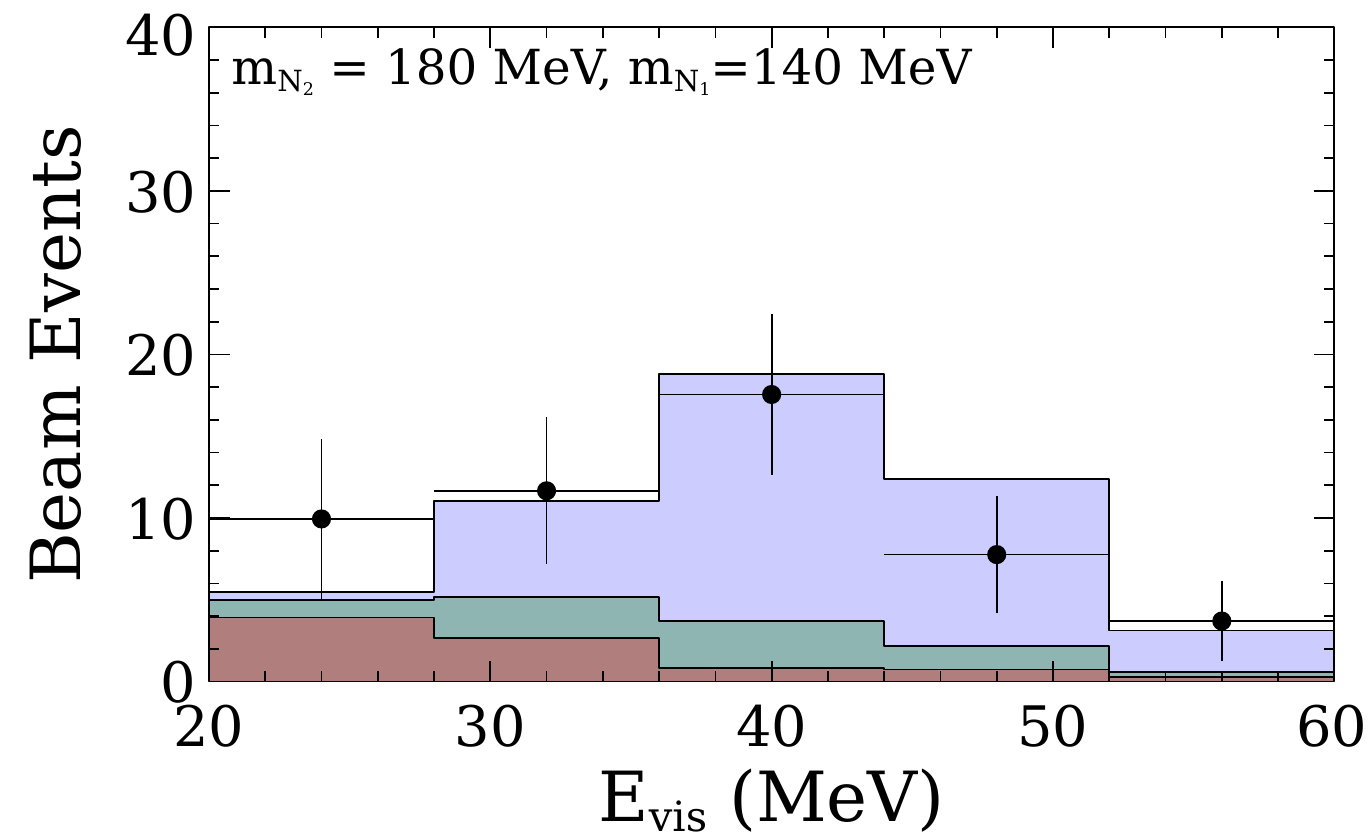}
\caption{The left (right) column corresponds to the MB (LSND) events as a function of the visible energy, $E_{\rm vis}$. Various values of the masses of the daughter particle, $m_{N_1}$ are shown in the plot. The mass of the single scalar is fixed at 300 MeV  for all panels while the mass of  the decaying particle $N_2$ is assumed  to be 135 MeV for the top and middle panels and 180 MeV in the bottom panel.}
\label{energy-dist}
 \end{center}
\end{figure*}

 Proceeding on the basis of what we have learnt with respect to requirements in the previous sections,  we next examine if  the observed energy distributions in LSND and MB can provide additional information. In principle, one has  the freedom to choose the masses of the mediator, $N_1$ and $N_2$.  Recalling that  $m_{N_2}\sim 100$~MeV and a single 300~MeV scalar mediator appear to be reasonable choices in our search for acceptable benchmarks and putting  the constraint considerations mentioned above which restrict a $300$ MeV singlet scalar aside for the moment, we examine its effect on the energy distributions. As above, we  assume  that $N_2$ once produced, decays to a second lighter heavy sterile neutrino $N_1$ (or $\nu_i$) and either a $\gamma$ or an $e^+e^-$ pair.  Figure~\ref{energy-dist} shows the distributions for MB and LSND for three different choices of the $N_1$ mass (60~MeV, 100~MeV, 140~MeV), two different mass values of $N_2$ (135~MeV in the first two rows and 180~MeV in the bottom row) and  a fixed mass value of $300$~MeV for the scalar mediator. We note that the scalar mediates both the interaction that produces $N_2$ and the subsequent decay leading to the signal.
 
Considered  in conjunction, the panels in figure~\ref{energy-dist} reflect the difficulties in obtaining a good fit simultaneously for the MB and LSND energy distributions if a single scalar is used. The top panels, where a relatively low mass of $N_2$ is assumed, provides a reasonable fit to the MB distribution. On the other hand, while not apparent in the figure,  the low mass  pushes the  LSND distribution to the right, leading  to too many events with energies greater than $60$~MeV, which were not seen. On increasing the $N_2$ mass moderately, to $100$~MeV (middle panels) one finds that while the LSND distribution appears to have a  shape closer to that which was observed, the total number of events is much higher than required. The MB fit worsens compared to the top panel, with events being underproduced. It is possible to reduce the total number of events in LSND by increasing the mass of $N_2$, which is what we attempt to do in the bottom panels. This certainly improves the LSND fit, but the MB distribution, although shifted, continues to significantly underproduce  events in many bins, especially  in the 250~MeV to 550~MeV range. In general, attempts to fit LSND will cause a depletion in the low and/or medium energy bins of MB, which contain the bulk of its event excess. For example, the total number of MB events in the bottom left panel is 294  compared to the observed value of 560.  
 We also note the reason for the overproduction of events in many cases in LSND:  this is because cuts which ensure that the pairs  produced in MB mimic the signal of a single charged lepton have been applied. These consist in requiring a)~that the opening angle between the pair is $\leq 10^\circ$ or, b)~in the case of an asymmetric pair, the softer partner carries an energy $\leq 30$~MeV. Once this is done, only about $20-30\%$ of total  MB events  survive. This requires us to have a higher coupling when fitting MB, which, when used to calculate the LSND events, where no cuts are applied, leads to a corresponding higher number\footnote{The same behaviour is expected in the case of a vector mediator. However, as we have discussed earlier, vector mediators are subject to strong cross section constraints.}. This behaviour is aided by the tendency of the incoherent scalar mediated cross section to rise as the energy is lowered from MB to LSND values, as clearly seen from figure~\ref{x1} (left panel).
 
 \section{Requirements resulting from considerations related to the visible signal in LSND and MB}
 \label{sec7}
 A detailed study of the requirements stemming from the observed signal (figures~\ref{FD}(b)~and~\ref{FD}(c)) depends on specifics of the model employed, which is beyond the scope of this work. Hence in this section we restrict ourselves to certain general observations as well as point to constraints which must be kept in mind.
 
The simplest possibilities which could mimic  an electron-like signal in new physics scenarios are  either a  $\gamma$ or a collimated $e^+ e^-$ pair, which, in the scenario considered here, result from the decay of $N_2$, generically governed by one of the following interaction Lagrangians:
\begin{eqnarray}
{\cal L}_{\rm int}&\supset& \lambda^H_{N_{12}} \bar{N}_1 N_2 H+y_e^H \,\bar{e}\, e\, H,\\
{\cal L}_{\rm int}&\supset& \mu_{\rm tr}\, \bar\nu_{i L}\,\sigma_{\mu\nu}\,F^{\mu\nu} N_{R2},
\end{eqnarray}
where $\lambda^H_{N_{12}}$ and $y_e^H$ are Yukawa couplings. $\mu_{\rm tr}$ is the neutrino transition magnetic moment, $F^{\mu\nu}$ is the electromagnetic field strength tensor and $\sigma_{\mu\nu}=\frac{i}{2}[\gamma_\mu,\gamma_\nu]$.

From kinematics, $N_2$ ($\sim 150$~MeV) cannot decay to an on-shell scalar of mass $\sim 300$ MeV, which are the preferred ball-park mass values we have identified for the heavy sterile neutrino and the mediator respectively  from the discussions in previous sections. The decay process could happen via  the off-shell exchange of the scalar, with  $N_2 \rightarrow N_1 + e^+ + e^-$. A discussion of constraints related  to the production of an  $e^+ e^-$ pair in the final state is provided in~\cite{MiniBooNEDM:2018cxm,Arguelles:2018mtc,1810.07185,Brdar:2020tle, Abdallah:2020vgg}. 
 
 In general, the kinematics of  three-body decay allows the pair to have an energy greater than $\frac{1}{2} m_{N_2}$ in the rest frame of $N_2$ (unlike  two-body decay). This is all the more possible if  $N_1$ is relatively light, $\ie$ $m_{N_1}\leq 60$~MeV, or if one were to consider the signal as arising from  $N_2 \rightarrow \nu_i + e^+ + e^-$, where $\nu_i$ is an active SM neutrino. Overall, because of this, the three-body decay option which one is forced into when the mediator mass exceeds that of $N_2$ thus tends to give too many excess events in LSND beyond visible energies of $60$~MeV, in conflict with observations. This last consideration also holds for the case where the excess results from  $N_2 \rightarrow \nu_i + \gamma$, mediated by a transition magnetic moment\footnote{The process $N_2\to N_1+\gamma$ is also possible. In the simplest case this would proceed via mixing with SM neutrinos but would be highly suppressed. Other possibilities would be model dependent, for example, via a charged Higgs in the model discussed in~\cite{Abdallah:2020vgg}.}.
 Depending on the model details, solutions where the signal arises due to a photon coupling via a magnetic moment must contend with strong constraints on either active-sterile mixing
 and/or magnetic moments~\cite{0901.3589,1011.3046,1110.1610,Corsico:2014mpa,1502.00477,1511.00683,Magill:2018jla,1904.06787,1909.11198,Diaz:2019kim,Studenikin:2021fmn} or from the requirement to match  the observed angular distribution in MB~\cite{Radionov:2013mca}.
 
 An additional important restriction for the three-body decay situation arises from mono-photon searches at BaBar~\cite{Lees:2017lec}. These  put a tight constraint on a dark photon ($A'$) that decays invisibly. The bound on the kinetic mixing parameter $\varepsilon$ is  $(e\varepsilon)^2<9\times10^{-8}$ for $m_{A'} \lesssim 6$~GeV coming from the process $e^+e^- \to\gamma A'$; $A' \rightarrow$ invisibles, where $e\simeq 0.3$ is the electric charge. This strong limit on $\varepsilon$ can be relaxed if the $A'$ decays semi-visibly inside the detector \textit{i.e.} BR$(A'\to {\rm visible~particle~+~missing~energy})\simeq 1$~\cite{Mohlabeng:2019vrz,Duerr:2019dmv}.  For our purpose here, the light scalar ($H$) will be produced along with a  $\gamma$ in $e^+e^-$ collision. Since  $H$ also decays semi-visibly via $H \rightarrow N_1 + N_2 \rightarrow N_1+N_1+e^+ +e^-$ with almost $100\%$ branching ratio,  the resultant  constraint on $y^H_e$ can be relaxed so that $N_2$ decays promptly within the LSND and MB detectors. Similar considerations were  used in Ref.~\cite{Abdullahi:2020nyr}. Our calculations here use an  effective coupling ($y^H_e \lambda^{H}_{N_{12}}$) of $9\times 10^{-4}$ to enable this,  which  gives a decay length of ${\cal O}(\sim$ few cm) for $N_2$.
 \section{Discussion}
 \label{sec8}  
  Our examination of the cross section behaviour with energy  and its dependance on the mediator mass (section~\ref{sec3}) suggests that a common solution is simpler to achieve  using scalar mediators as opposed to vectors. Vector solutions which are likely to provide reasonable fits to the angular and energy distributions in MB tend to give lower cross sections than required when it comes to LSND, producing too few events once the new physics parameters are fixed by a MB fit. Such  mediators also have cross sections which stay flat  in this energy range as the neutrino energy increases, which makes it difficult to satisfy constraints from higher energy experiments like CHARM~II~\cite{CHARM-II:1994dzw}, MINER$\nu$A~\cite{Valencia:2019mkf} and IceCube~\cite{Coloma:2017ppo,Coloma:2019qqj}.
  
 Two additional considerations which apply to vector solutions are:
 \begin{itemize} 
 \item In their  search for a final state photon produced by $\Delta$ production and decay~\cite{MicroBooNE:2021zai}, MicroBooNE found no evidence of an excess. In models with an additional vector which can mediate the interaction, an additional contribution to $\Delta$ production is expected to  occur, rendering them potentially in tension with  this observation if this contribution is statistically significant. 
 \item  Any new light vector with dimension-4 couplings to the SM fermions must couple to a conserved current. If this is not the case,  processes with (energy/vector mass)$^2$ rates involving the longitudinal mode of the new vector  can be the dominant production mechanism in high-energy experiments, which places strong constraints on its coupling~\cite{Kahn:2016vjr,1705.06726,1707.01503}.
 \end{itemize}
  Considerations based on the observed angular distributions in LSND and MB are discussed in section~\ref{sec5}. The experimentally observed  distribution for MB is fairly forward, but the excess extends in all directions, although it becomes smaller as we move away from the $\cos\theta\simeq 1$ values. As expected, lighter mediators, whether scalar or vector, favour low momentum transfers to the target and hence tend to populate the forward direction more so than heavier ones. As the mediator mass is increased, other angular bins get populated, while the forward direction, which is dominantly populated in the observed distribution, gets depleted. We find that an intermediate mass mediator does offer the possibility of a reasonable  fit to the observed angular features in MB. However, considerations based on the observed energy distributions in the two experiments, which we have examined in section~\ref{sec6} above, disfavour a single  intermediate mass scalar mediator, because it becomes difficult to simultaneously find good fits to the observed energy distributions in LSND and MB.
  
  Thus, overall, energy distributions in LSND and MB, the angular distribution in MB, when combined with the stringent constraints on light singlet scalars, suggest the use of a scalar doublet, with one light and one moderately heavy partner.  This allows for a  degree of angular isotropy as well as  a large number of events in the forward direction, as were observed.
  With respect to energy distributions, the lighter partner tends to help give the correct distribution in LSND, while heavier partners tend to distort it. As one moves from MB energies down to LSND energies, the incoherent cross section  (relevant for LSND) for heavier mediators drops much more rapidly than lighter mediators, which consequently contribute more to the total events. Thus, a combination of a moderately heavy and a light mediator complement each other  well when a common solution to the two anomalies is sought.   An example solution  to both  anomalies   that incorporates  all the features that have emerged in our study has been provided in~\cite{Abdallah:2020vgg}.
  
 Our approach has been phenomenological, keeping simplicity and minimality as guiding principles to the extent possible.  Our results should be treated as indicative, and not definitive, since it will  always be possible to construct models that circumvent specific obstacles. Such efforts would, however, extract a price in terms of increased complexity  in  the number of particles and new interactions invoked. We stress that the requirements we obtain above  pertain to {\it{simultaneous}} solutions  which attempt to explain both the LSND and MB anomalies, and can be considerably different if this condition is relaxed.
\section{Conclusions} 
\label{sec9} 
In this work, we have focussed on an important class of common, non-oscillation new physics solutions to the LSND and MB excesses, keeping the recent MicroBooNE results in mind. It comprises of solutions which have an $e^+e^-$ pair or a photon produced inside the detector via a new interaction involving the up-scattering of a heavy sterile neutrino and its subsequent decay.  We have used cross section magnitudes and variation with energy and angle as well as existing constraints  to arrive at conclusions which point to scalar mediators being favoured over vectors, and to one light ($\lesssim50$~MeV) and one moderately heavy ($\lesssim 850$~MeV) mass mediator being a better choice than a single scalar of intermediate mass. 

The insistence on a solution that resolves both excesses simultaneously is, of course, a choice. It restricts and guides potential solutions in ways that attempts to address the anomalies individually do not. It is, however, remarkable that  once this is demanded, and the dictates of the cross section,  the observed energy and angular distributions in both experiments  as well as the many constraints from various experiments adhered to (this work), then one is led to a simple extension of the SM that i) resolves both anomalies, ii) provides a portal to the dark sector, iii) accounts for the experimentally observed value of the muon $g-2$ and iv) addresses the issue of neutrino mass via a Type I seesaw, in conformity with the observed values of neutrino mass-squared differences in oscillation experiments, as shown in~\cite{Abdallah:2020vgg}.

Additionally, we have made an effort to emphasise that understanding the LSND and MB results in conjunction with MicroBooNE provides both a high-stakes challenge as well as an unprecedented opportunity for the discovery of physics beyond the SM. 
 An oft-repeated statement in the literature is that the only  physics  discovery beyond the SM, since its inception over fifty years ago is that neutrinos are massive. While the theoretical, phenomenological  and experimental consequences of neutrino mass have been profound, altering  our approach to  and understanding of neutrino physics, cosmology, stellar evolution, dark matter, leptogenesis and CP violation, it is also true that the {\it{structural}} change in the SM  necessary  to accommodate it is relatively benign, requiring the addition of right-handed neutrinos only\footnote{This is true irrespective of whether they are Majorana or Dirac particles, albeit the mass scales at which these neutrinos enter the theory can be very different for the two.}. If indeed, the present results that are the subject of this paper do herald new physics, this is likely {\it{not}} to be the case.  All of the proposals for the three possible final states  (single electrons, an $e^+e^-$ pair or a photon) require non-trivial new physics. In particular, those giving  an $e^+e^-$ pair do not have a known SM background, and appear to call for  new particles and interactions, as well as some portal to the dark sector, indicating possibly  major additions to the particle spectrum and the interactions between them. 

It is fortuitous that all three types of possible final states will be tested soon, first in MicroBooNE and then in the other detectors comprising the Fermilab short baseline program. \\ 
The anticipation is palpable.
\acknowledgments
 RG is  grateful to William Louis for his help with our many questions on LSND, MB and MicroBooNE. He would also like to express appreciation of the many interesting conversations he has had with Boris Kayser and Geralyn Zeller on the anomalies which are the subject of this paper.
 
 WA, RG and SR also acknowledge support from the XII Plan Neutrino Project of the Department of Atomic Energy and the High Performance Cluster Facility at HRI (http://www.hri.res.in/cluster/).
\bibliographystyle{apsrev}
\bibliography{NU-bib}
\end{document}